\documentclass[%
prx
,twocolumn%
,amsmath,amssymb,aps,nobibnotes
,notitlepage
,nofootinbib
,preprintnumbers
]{revtex4-1}

\usepackage[colorlinks,linkcolor=blue,citecolor=blue]{hyperref}
\usepackage[dvipdfmx]{graphicx}
\usepackage{bm}
\usepackage{braket}
\usepackage{color}
\usepackage{tabularx}
\usepackage{mathtools}

\begin{document}
\title{Cavity Moir\'e Materials: Controlling magnetic frustration with quantum light-matter interaction}
\author{Kanta Masuki}
\email{masuki@g.ecc.u-tokyo.ac.jp}
\affiliation{Department of Physics, University of Tokyo, 7-3-1 Hongo, Bunkyo-ku, Tokyo 113-0033, Japan}

\author{Yuto Ashida}
\email{ashida@phys.s.u-tokyo.ac.jp}
\affiliation{Department of Physics, University of Tokyo, 7-3-1 Hongo, Bunkyo-ku, Tokyo 113-0033, Japan}
\affiliation{Institute for Physics of Intelligence, University of Tokyo, 7-3-1 Hongo, Tokyo 113-0033, Japan}

\begin{abstract}
	Cavity quantum electrodynamics (QED) studies the interaction between light and matter at the single quantum level and has played a central role in quantum science and technology. Combining the idea of cavity QED with moir\'e materials, we theoretically show that strong quantum light-matter interaction provides a way to control frustrated magnetism. Specifically, we develop a theory of moir\'e materials confined in a cavity consisting of thin polar van der Waals crystals. We show that nontrivial quantum geometry of moir\'e flat bands leads to electromagnetic vacuum dressing of electrons, which produces appreciable changes in single-electron energies and manifests itself as long-range electron hoppings. We apply our general formulation to a twisted transition metal dichalcogenide heterobilayer encapsulated by ultrathin hexagonal boron nitride layers and predict its phase diagram at different twist angles and light-matter coupling strengths. Our results indicate that the cavity confinement enables one to control magnetic frustration of moir\'e materials and might allow for realizing various exotic phases such as a quantum spin liquid.
\end{abstract}

\maketitle

\section{Introduction\label{sec:intro}}
Controlling exotic phases of matter has been an ongoing quest in condensed matter physics. Moir\'e materials are emerging platforms for studying strongly correlated phenomena, where a long-periodic moir\'e pattern is formed by two overlaid crystal layers with relative twists or different lattice constants. Such moir\'e superlattice induces reconstruction of electronic structures and realizes nearly flat bands at certain twist angles~\cite{lopesdossantos2007Graphene,bistritzer2011Moire,carr2020Electronicstructure}.
In flat-band systems, the kinetic energy of electrons is significantly suppressed and the effect of electron-electron interaction becomes important~\cite{KM18,PHC18}. So far, a number of intriguing phenomena have been experimentally observed in twisted bilayer graphene~\cite{andrei2020Graphene,balents2020Superconductivity,nuckolls2020Strongly}, including metal-insulator transition~\cite{cao2018Correlated,burg2019Correlated,chen2019Evidence}, flat-band superconductivity~\cite{cao2018Unconventional,chen2019Signatures,codecido2019Correlated,lu2019Superconductors,yankowitz2019Tuning}, magnetism~\cite{sharpe2019Emergent,chen2020Tunable,liu2020Tunable,he2021Competing}, and fractional quantum Hall effect~\cite{dean2013Hofstadter,hunt2013Massive,wang2015Evidence,spanton2018Observation}.
Owing to rapid advances in manipulation of van der Waals (vdW) heterostructures~\cite{liu2019Van,vincent2021Opportunities}, moir\'e materials consisting of transition metal dichalcogenides (TMDs) have also been extensively investigated~\cite{kennes2021Moire,wu2018Hubbard,classen2019Competing,xian2019Multiflat,WM2019-2,pan2020Band,pan2020Quantum,angeli2021Valley,hu2021Competing,morales-duran2021Metalinsulator,zang2021HartreeFock,xian2021Realization,zang2022Dynamical}.
In particular, the high tunability of TMDs enables one to study various correlated phases~\cite{jin2019Observation,ni2019Soliton,regan2020Mott,shimazaki2020Strongly,tang2020Simulation,wang2020Correlated,xu2020Correlated,zhang2020Flat,huang2021Correlated,jin2021Stripe,yasuda2021Stackingengineered,li2021Continuous} and might allow for realizing exotic states such as a quantum spin liquid (QSL)~\cite{savary2016Quantum,zhou2017Quantum,zare2021Spin,kiese2022TMDs}.

On another front, experimental developments in cavity quantum electrodynamics (QED) have allowed for realizing the ultrastrong coupling regime, where light-matter interaction is comparable to elementary excitation energies \cite{anappara2009Signatures,scalari2012Ultrastrong,scalari2014,maissen2014Ultrastrong,gambino2014Exploring,chikkaraddy2016Singlemolecule,yoshihara2017Superconducting,bayer2017Terahertz,halbhuber2020Nonadiabatic,genco2018Bright,flick2017Atoms,forn-diaz2019Ultrastrong,friskkockum2019Ultrastrong}.
Recent studies have discussed the cavity confinement as an alternative way to control the phase of matter without an external drive in a wide variety of fields, including quantum optics \cite{hubener2021Engineering,ruggenthaler2018,flick2019,mueller2020Deep,owens2022,ashida2022Nonperturbative}, polaritonic chemistry \cite{ebbesen2016,feist2018,ribeiro2018,galego2015}, and condensed matter physics \cite{garcia-vidal2021Manipulating,schlawin2022Cavity,bloch2022Strongly,orgiu2015,konishi2021,pilar2020Thermodynamics,roman-roche_effective_2022,ZZ21,juraschek2021Cavity,lenk2022Collective,ashida2020Quantum,latini2021Ferroelectric}.
In particular, the possibility of controlling certain material properties, such as topological aspects~\cite{mann2018Manipulating,wang2019,rokaj2019Quantum,*rokaj2022Polaritonic,downing2019Topological,tokatly2021,MK22}, superconductivity~\cite{sentef2018Cavity,schlawin2019CavityMediated,curtis2019Cavity,curtis2022Cavity,li2022}, correlated phenomena \cite{thomas2021,kiffner2019Manipulating,*kiffner2019Mott,chiocchetta2021Cavityinduced,li2020,VB22}, and Landau polaritons~\cite{smolka2014Cavity,zhang2016Collective,paravicini-bagliani2019Magnetotransport,keller2020,appugliese2022Breakdown,ravets2018,hagenmuller2010Ultrastrong}, has been so far explored.
On the one hand, due to the smallness of the fine structure constant, the single-quantum ultrastrong coupling is out of reach in a usual Fabry-Perot cavity consisting of metallic mirrors \cite{DMH07}. On the other hand, this difficulty can be overcome by employing hybridization with matter excitations. For instance, a recent study  \cite{YA23} has shown that a new cavity geometry, in which two ultrathin vdW layers form a planar cavity, can achieve single-quantum ultrastrong couplings and thus provides an ideal platform to explore ultrastrong coupling physics of two-dimensional electronic materials. There, an optical anisotropy in vdW layers leads to the formation of phonon polaritons having hyperbolic dispersion \cite{JZ14}. The electrons are then strongly coupled to tightly confined hyperbolic polaritons, where the coupling strength can be tuned simply by changing thicknesses of vdW slabs. Given these developments and prospects, it is natural to address whether or not the cavity confinement enables one to control correlated phases of moir\'e materials.

\begin{figure}[b]
	\includegraphics[width = 8.6cm]{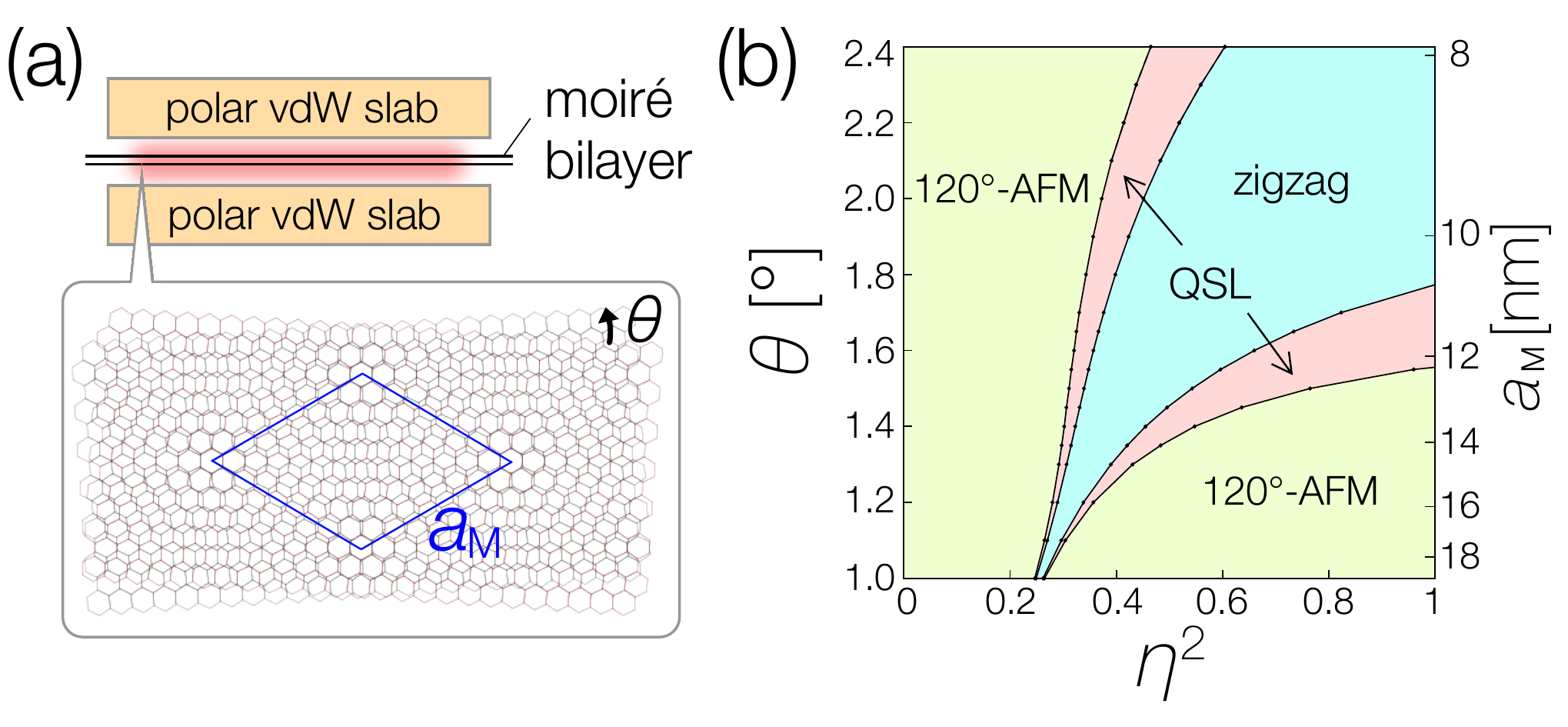}
	\caption{\label{figure1}(a) Schematic figure of cavity moir\'e materials. Electrons in a moir\'e bilayer are ultrastrongly coupled to hyperbolic phonon polaritons confined in the cavity consisting of polar vdW crystals such as \textit{h}-BN. The bottom panel shows a moir\'e superlattice with lattice constant \(a_M\) formed by a bilayer with twist angle \(\theta\). (b) Ground-state phase diagram of the cavity-confined TMD heterobilayer at half-filling plotted against \(\theta\) and dimensionless light-matter coupling strength \(\eta\). Tuning \(\eta\) and \(\theta\), one can control various correlated phases, such as the 120\(^\circ\)-antiferromagnetic (AFM) phase, the zigzag phase, and a candidate quantum spin liquid (QSL) phase.}
\end{figure}

In this paper, taking a step beyond Ref.~\cite{YA23}, we show that strong quantum light-matter interaction provides a way to control many-body properties of moir\'e materials. Specifically, we develop a theory of cavity moir\'e materials to describe the interaction between electrons in a twisted bilayer and quantized electromagnetic fields inside a cavity (Fig.~\ref{figure1}(a)) and show that magnetic frustration in moir\'e materials can be controlled by the cavity confinement.
A key point is that, unlike isolated flat-band systems, moir\'e materials are inherently multiband systems with large interlayer contributions in tensor Berry connections \cite{torma2022Superconductivity,topp2021Lightmatter}.
We show that this quantum geometric effect causes vacuum-induced virtual electronic transitions between flat bands and the other bands, leading to the renormalization of single-electron energies in flat bands. While such vacuum-induced modification is usually irrelevant in conventional materials, we point out that it becomes important in moir\'e materials because of their extremely narrow bandwidths.

We demonstrate that this electromagnetic vacuum dressing allows for controlling a variety of correlated phases of moir\'e materials, some of which remain elusive in current experiments. Specifically, we apply our theory to a TMD moir\'e heterobilayer encapsulated by ultrathin hexagonal boron nitride (\textit{h}-BN) slabs (Fig.~\ref{figure1}(a)), where light-matter coupling strength $\eta$ can be tuned by varying \textit{h}-BN thicknesses. At small twist angle \(\theta\) and half-filling, low-energy physics of the flat-band electrons can be described by the spin-$1/2$ antiferromagnetic (AFM) Heisenberg model on the triangular moir\'e lattice~\cite{wu2018Hubbard,downing2019Topological,macdonald1988FractU}.
When placed inside the cavity, vacuum fluctuations appreciably renormalize the single-electron energies and induce long-range electron hoppings.
As a result, the cavity confinement enhances the spin frustration in the Heisenberg model and allows one to control various phases including the 120\(^\circ\)-AFM phase and the zigzag phase (Fig.~\ref{figure1}(b)). Notably, in the intermediate regions, one may even realize a QSL phase of the triangular Heisenberg model, whose nature has been recently under debate \cite{ADP22,zhu2015Spin,gong2019Chirala,hu2019Dirac,szasz2020Chiral,drescher2022Dynamical}. We expect that these predictions are within experimental reach owing to recent developments demonstrating ultrasmall mode volumes of hyperbolic polaritons in nanostructured materials \cite{CJD14,SD14,GA18,SD19,MEY22,SHH22}.

The rest of this paper is organized as follows. In Sec.~\ref{sec:Model}, we introduce the total Hamiltonian to describe moir\'e materials confined in hyperbolic cavities.  In Sec.~\ref{sec:Eff_Ham}, we derive the low-energy effective Hamiltonian of flat-band electrons by performing the perturbative analysis, which is one of the main results of this paper. Importantly, the effect of the cavity confinement is represented by the dressing of single-electron energies of the flat band. In Sec.~\ref{sec:general}, we examine the general properties of the effective Hamiltonian and show that the energy-dressing of flat-band electrons enhances the long-range couplings in real space, whose characteristic length scale can be controlled by  the light-matter coupling strength and the twist angle of moir\'e materials. Lastly, in Sec.~\ref{sec:spin-model}, we employ our formalism to analyze the spin-ground state of flat-band electrons in WSe\(_2\)/MoSe\(_2\). Our results indicate that the cavity confinement enables one to control the magnetic frustration of moir\'e materials and might allow for realizing various many-body phases. Section~\ref{sec:Discussion} summarizes the results and discusses the future directions for nonperturbative analysis of cavity moir\'e materials. Some technical details are discussed in Appendix.

\section{Model Description\label{sec:Model}}

\begin{figure}[t]
	\includegraphics[width = 8.6cm]{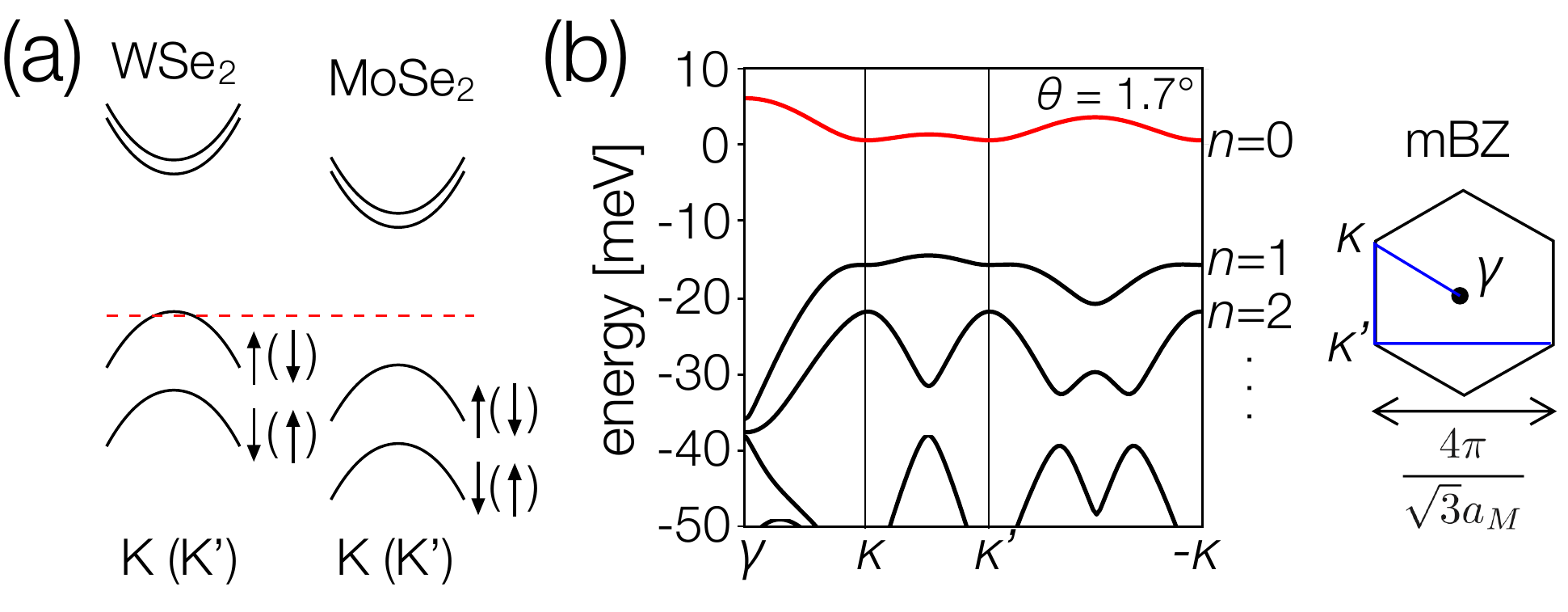}
	\caption{\label{figure2} (a) Schematic band dispersions of monolayer WSe\(_2\) and MoSe\(_2\). The valence band maxima (VBM) of WSe\(_2\) at different valleys with the opposite spin are located in the band gap of MoSe\(_2\). The red dashed line represents the Fermi level. (b)  Moir\'e bands of Eq.~\eqref{moi_hamiltonian} at \(\theta = 1.7^\circ\). The nearly flat band with $n=0$ is highlighted by red color. The right panel shows representative points on the moir\'e Brillouin zone (mBZ). Parameters are \((w,\psi) = (6.6\ {\rm meV}, -94^\circ)\) and \(m^* = 0.35m_e\), where \(m_e\) is the electron mass~\cite{wu2018Hubbard}.}
\end{figure}

To be concrete, we focus on a cavity confinement of twisted TMD heterobilayer WSe\(_2\)/MoSe\(_2\), while our theoretical formulation can be generally applied to other moir\'e materials (see Appendix~\ref{sec:app:derivation} for details). Due to the spin-orbit coupling, the valence bands of each monolayer have two band maxima at different valleys with opposite spin degrees of freedom (Fig.~\ref{figure2}(a))~\cite{xiao2012Coupled,wu2018Hubbard}. Since the valence band maxima (VBM) of monolayer WSe\(_2\) are located in the band gap of MoSe\(_2\), we can analyze the bilayer in terms of the VBM electrons in WSe\(_2\) provided that the chemical potential is tuned near the VBM of WSe\(_2\) (red dashed line in Fig.~\ref{figure2}(a))~\cite{wu2018Hubbard, carr2020Electronicstructure,zhang2016Systematic}.

When the TMD bilayer is twisted with small angle \(\theta\), a triangular moir\'e superlattice with lattice constant \(a_M\!\approx\! a_0/\theta\) is formed (see Fig.~\ref{figure1}(a)), where \(a_0\!\simeq\!3.32\) \AA\ is the monolayer lattice constant of WSe\(_2\). The effect of the moir\'e superlattice can be described by an effective single-particle potential \(\Delta(\bm r)\), which has the same periodicity as the moir\'e superlattice~\cite{wu2018Hubbard,carr2020Electronicstructure}. The low-energy Hamiltonian of the VBM electrons thus reads
\begin{align}
	\hat H_{0} & = - \frac{\hat{\bm p}^2}{2m^*} + \Delta(\hat{\bm r})\label{moi_hamiltonian},
\end{align}
where \(m^*\) is the effective mass in the valence band of WSe\(_2\), and the moir\'e potential \(\Delta(\bm r)\) is given by
\begin{align}
	\Delta(\bm r) = \sum_{j=1}^6 w_j e^{i\bm b_j\bm r}.\label{moi_potential}
\end{align}
Here, \(\bm b_j\) is the reciprocal lattice vector of the moir\'e superlattice corresponding to \((j-1)\pi/3\)-rotation of \(4\pi/(\sqrt{3}a_M)\bm e_x\). Since \(\Delta(\bm r)\) is real-valued and satisfies the threefold-rotational symmetry, the coefficients \(w_j\) can be parametrized as \(w_j = we^{-i(-1)^j\psi}\). 

In Fig.~\ref{figure2}(b), we show typical band dispersions \(\varepsilon_{n\bm k}\) of the moir\'e Hamiltonian~\eqref{moi_hamiltonian} on the moir\'e Brillouin Zone (mBZ), where one can see the nearly flat band \((n\!=\!0)\) located above the other bands \((n\!\geq\!1)\). It is noteworthy that, due to the nontrivial quantum geometry of the moir\'e flat band, the electron momentum \(\hat{\bm p}\) has nonvanishing interband matrix elements bewteen the flat band and the other bands below, while the innerband matrix element of \(\hat{\bm p}\) within the flat band almost vanishes due to the band flatness \cite{torma2022Superconductivity,topp2021Lightmatter}. Since electrons couple to the dynamical electromagnetic fields with vector potential \({\bm A}\) through \(\hat{\bm p}\!\cdot\!{\bm A}\), it is these nonvanishing matrix elements that facilitate the couplings between flat-band electrons and cavity fields as detailed below.

The total Hamiltonian of the cavity-confined bilayer is given by minimally coupling the vector potential operator of cavity fields \(\bm \hat {\bm A}\) to the momentum operator of electrons \(\hat{\bm p}\). In the second-quantized form, the total Hamiltonian is expressed as 
\begin{align}
	\hat H              = & \sum_{\sigma}\int\!d^2\bm r\ \hat \psi_{\sigma\bm r}^\dagger \left[ -\frac{(-i\hbar\nabla +e\hat{\bm A}(\bm r))^2}{2m^*}\!+\!\Delta(\bm r) \right] \hat \psi_{\sigma\bm r}\nonumber \\
	                      & +\sum_{\bm q} \hbar\omega_{\bm q} \hat a_{\bm q}^\dagger \hat a_{\bm q}+\hat U,\label{cav_moi_ham1}
\end{align}
where we also include the Coulomb interaction \(\hat U\). 
Here, \(\hat\psi_{\sigma\bm r}\) (\(\hat \psi_{\sigma\bm r}^\dagger\)) is the annihilation (creation) operator of electrons with spin \(\sigma\) at position \(\bm r\), \(\hat a_{\bm q}\) (\(\hat a_{\bm q}^\dagger\)) is the annihilation (creation) operator of hyperbolic polaritons with in-plane momentum $\bm q$, and $\omega_{\bm q}$ is the mode frequency. The vector potential \(\hat{\bm A}(\bm r)\) is
\begin{align}
	\hat{\bm A} (\bm r) = \sum_{\bm q}A_{\bm q}\bm e_{\bm q}(\hat a_{\bm q}e^{i\bm q\bm r}+\hat a_{\bm q}^\dagger e^{-i\bm q\bm r}),\label{vec_pot_operator}
\end{align}
where $A_{\bm q}$ is the mode amplitude of the electromagnetic component of polaritons, and \(\bm e_{\bm q}\!\equiv\!\bm q/|\bm q|\) is the effective polarization obtained after projecting the originally transverse 3D vector field onto the 2D plane where the bilayer is located.
We define the integrated dimensionless coupling strength \(\eta\) by 
\begin{align}
    \eta \equiv \sqrt{\sum_{\bm q}g_{\bm q}^2/\omega_{\bm q}^2},
\end{align}
where \(g_{\bm q}\) is the coupling strength between electrons and each cavity mode defined as
\begin{align}
    g_{\bm q} \!\equiv\! eA_{\bm q}\sqrt{\omega_{\bm q}/m^*\hbar}.
\end{align}
The value of \(\eta\) can reach up to $\eta\sim 1$ provided that thicknesses of \textit{h}-BN slabs are tuned to be a few nanometers \cite{YA23}. Lastly, we note that the phonon loss in vdW materials can in principle affect the mode amplitudes and the coupling strength of the hyperbolic phonon polaritons. However, as discribed in Appendix~\ref{sec:app:loss}, its effect is expected to be negligibly small in hyperbolic polaritons of \textit{h}-BN.

\section{Effective Hamiltonian of flat-band electrons\label{sec:Eff_Ham}}

\begin{figure}
    \includegraphics[width=8.6cm]{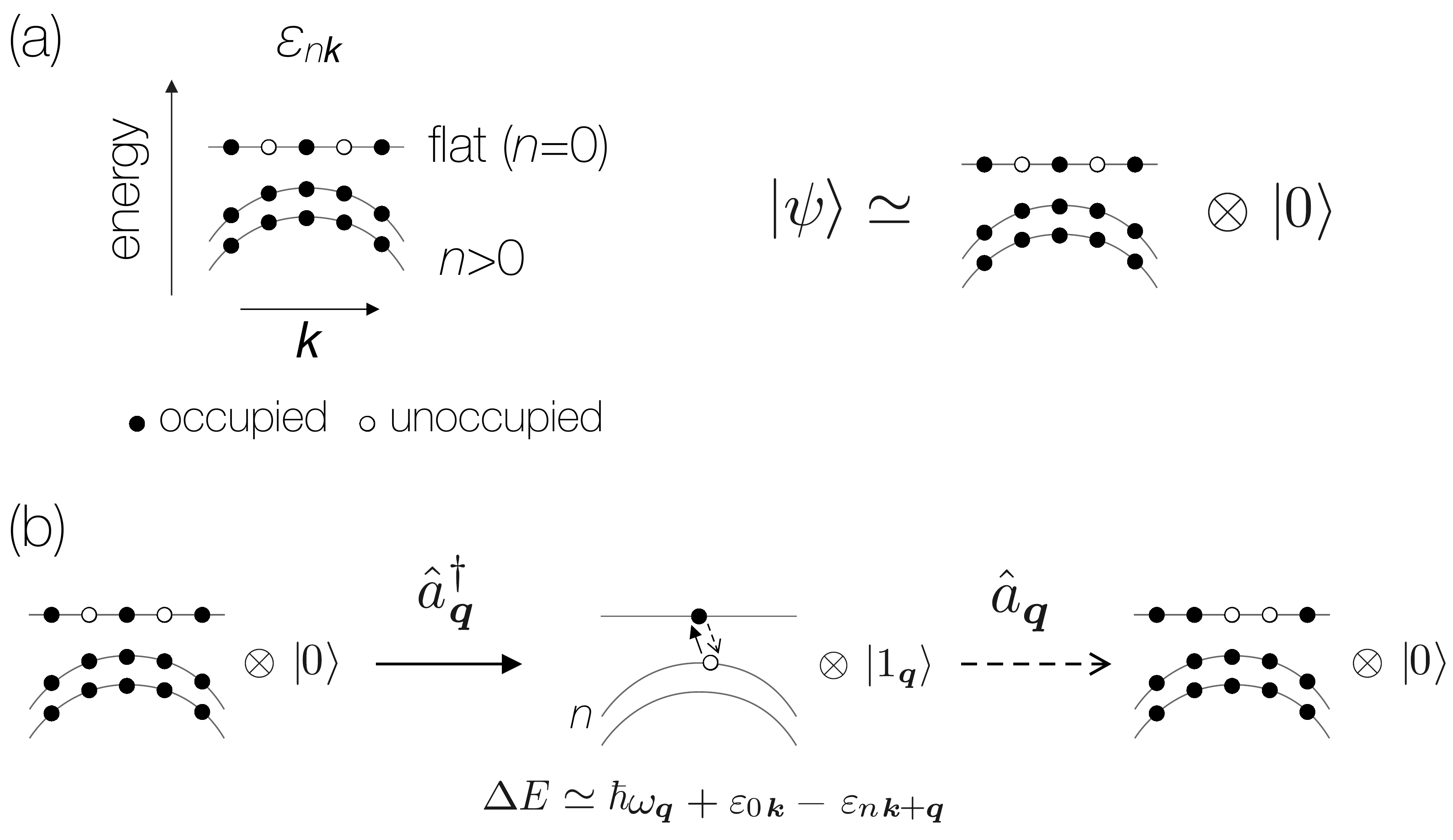}
    \caption{\label{fig:perturb_main} (a) Schematic band dispersions of the moir\'e Hamiltonian~\eqref{moi_hamiltonian}, where the flat band (\(n=0\)) is located above the fully occupied electronic bands. We approximate low-energy states \(|\psi\rangle \simeq |\psi_{\rm e}\rangle\otimes|0\rangle\) consisting of the product between the electromagnetic vacuum \(|0\rangle\) and an electronic state \(|\psi_{\rm e}\rangle\) with the partially-filled flat band and occupied lower bands. (b) Second-order-perturbation process that gives the last term in the right hand side of Eq.~\eqref{xi_k}. In this process, the electrons below the flat band and the cavity mode with in-plane momentum \(\bm q\) are simultaneously excited and then annihilated.}
\end{figure}

We are now in a position to derive the effective Hamiltonian of flat-band electrons. To this end, we first note that the cavity frequency, which is an order of optical phonon frequency in vdW crystals \(\hbar\omega_{\bm q}= \!10^2\!\sim\!10^3\) meV, is much larger than band gaps in the twisted bilayer (cf. Fig.~\ref{figure2}(b)). Therefore, the low-energy states of cavity moir\'e Hamiltonian~\eqref{cav_moi_ham1} can be approximated by a product state \(|\psi_e\rangle \otimes |0\rangle\), where \(|0\rangle\) is the electromagnetic vacuum, and \(|\psi_{\rm e}\rangle\) is an electronic state with the partially filled flat band and occupied lower electronic bands (cf. Fig.~\ref{fig:perturb_main}(a)). We can then adiabatically eliminate the cavity modes and include their vacuum fluctuations by performing the perturbative analysis with respect to $\eta$.

As detailed in Appendix~\ref{sec:app:derivation}, the leading contributions come from the second-order perturbation of the paramagnetic interaction term 
\begin{align}
    \hat H_{I}^{(p)}
    &= \sum_{\sigma}\int\!d^2\bm r\ \hat\psi_{\sigma{\bm r}}^\dagger 
    \left[ \frac{i\hbar e(\nabla\!\cdot\!\hat{\bm A}(\bm r)+\hat{\bm A}(\bm r)\!\cdot\!\nabla)}{2m^*} \right] \hat\psi_{\sigma\bm r}\\
    &= - \sum_{\bm q\bm k\sigma}\sum_{mn} \frac{eA_{\bm q}}{\hbar}(\bm G_{mn}^{\bm k,\bm q}\!\cdot\!\bm e_{\bm q})\ \hat c_{m\bm k+\bm q\sigma}^\dagger\hat c_{n\bm k\sigma}\hat a_{\bm q} + {\rm H.c.},
\end{align} 
where \(\hat c_{n\bm k\sigma}^{(\dagger)}\) is the annihilation (creation) operator of $n$-th band electrons with the Bloch wavevector $\bm k$ and spin $\sigma$. Also, we introduce the multiband coefficients \(\bm G_{mn}^{\bm k}\) by
\begin{align}
    \bm G_{mn}^{\bm k} \equiv \langle u_{m\bm k}|\nabla_{\bm k}\hat H_{0\bm k}|u_{n\bm k}\rangle,\label{G_mn_def}
\end{align}
where $|u_{n\bm k}\rangle$ is the Bloch eigenstate of \(\hat H_{0\bm k} \!\equiv\! e^{-i\bm k\hat{\bm r}}\hat H_0e^{i\bm k\hat{\bm r}}\) with band index $n$. Using simplifications that are valid in moir\'e flat-band systems, we obtain the effective Hamiltonian of cavity moir\'e materials, which is one of the main results in this paper (see Appendix~\ref{sec:app:derivation} for detailed derivations):
\begin{align}
	\hat H_{\rm eff} & = \sum_{\bm k\sigma}(\varepsilon_{0\bm k} + \eta^2\xi_{\bm k}) \hat c_{\bm k\sigma}^\dagger\hat c_{\bm k\sigma} + \hat U, \label{cav_moi_ham3}                        \\
	\xi_{\bm k}      & \equiv \frac{m^*}{2\hbar^2}\left(-|\bm G_{00}^{\bm k}|^2 + \sum_{n\geq 1} \frac{\hbar\omega_{\bm 0}}{\hbar\omega_{\bm 0} + \varepsilon_{0\bm k}-\varepsilon_{n\bm k}}
	|\bm G_{0n}^{\bm k}|^2\right).\label{xi_k}
\end{align}
Here, we abbreviate the creation (annihilation) operator of flat-band electrons \(\hat c_{0\bm k\sigma}^{(\dagger)}\) as \(\hat c_{\bm k\sigma}^{(\dagger)}\).
Remarkably, the effect of light-matter interactions is simply represented by the energy dressing \(\eta^2\xi_{\bm k}\) characterized by the multiband coefficients \(\bm G_{mn}^{\bm k}\). We note that the last term in the right hand side of Eq.~\eqref{xi_k} originates from the second-order perturbation process that simultaneously excites electrons below the flat band and cavity modes and then annihilates them (cf. Fig~\ref{fig:perturb_main}(b)). We note that the vacuum fluctuations in the cavity also mediate the Amperean electron-electron interaction in general. However, since its coefficients are proportional to \((\bm G_{00}^{\bm k,\bm q}\!\cdot\!\bm e_{\bm q})(\bm G_{00}^{\bm k',\bm q*}\!\cdot\!\bm e_{\bm q})\approx(\nabla_{\bm k}\varepsilon_{0\bm k}\!\cdot\!\bm e_{\bm q})(\nabla_{\bm k}\varepsilon_{0\bm k'}\!\cdot\!\bm e_{\bm q})\), the Amperean interaction is negligibly small compared to the Coulomb interaction \(\hat U\) (see Appendix~\ref{sec:app:derivation}).

From a quantum geometrical viewpoint, the multiband coefficients \(\bm G_{mn}^{\bm k}\)~\eqref{G_mn_def} can be expressed as
\begin{align}
	\bm G_{mn}^{\bm k} = \delta_{mn}\nabla_{\bm k}\varepsilon_{n\bm k} + i(\varepsilon_{m\bm k}-\varepsilon_{n\bm k}){\mathcal A}_{mn}(\bm k)
	\label{G_mn}
\end{align}
with \({\mathcal A}_{mn}(\bm k) = i\langle u_{m\bm k}|\nabla u_{n\bm k}\rangle\) being the tensor Berry connections. In moir\'e materials, ${\mathcal A}_{mn}$ has large interband contributions, while $\nabla_{\bm k}\varepsilon_{0\bm k}$  vanishes in the flat-band limit \cite{torma2022Superconductivity,topp2021Lightmatter}. Thus, the renormalization \(\xi_{\bm k}\) in Eq.~\eqref{xi_k} is mainly attributed to its second term originating from virtual interband transitions induced by the vacuum fluctuations in the cavity. Also, due to the flatness of the moir\'e electronic band, the dispersion of the renormalization \(\xi_{\bm k}\) in Eq.~\eqref{cav_moi_ham3} becomes larger than the original band dispersion \(\varepsilon_{0\bm k}\). As detailed in Sec.~\ref{sec:spin-model}, this feature of cavity moir\'e materials allows one to control ground-state magnetic properties of flat-band electrons.
We note that these key multiband processes in moir\'e materials cannot be captured in an oversimplified description of cavity materials, such as the Peierls substitution of the single-band tight-binding model. 

\section{General properties in cavity moir\'e materials\label{sec:general}}

\begin{figure}[t]
    \includegraphics[width = 8.6cm]{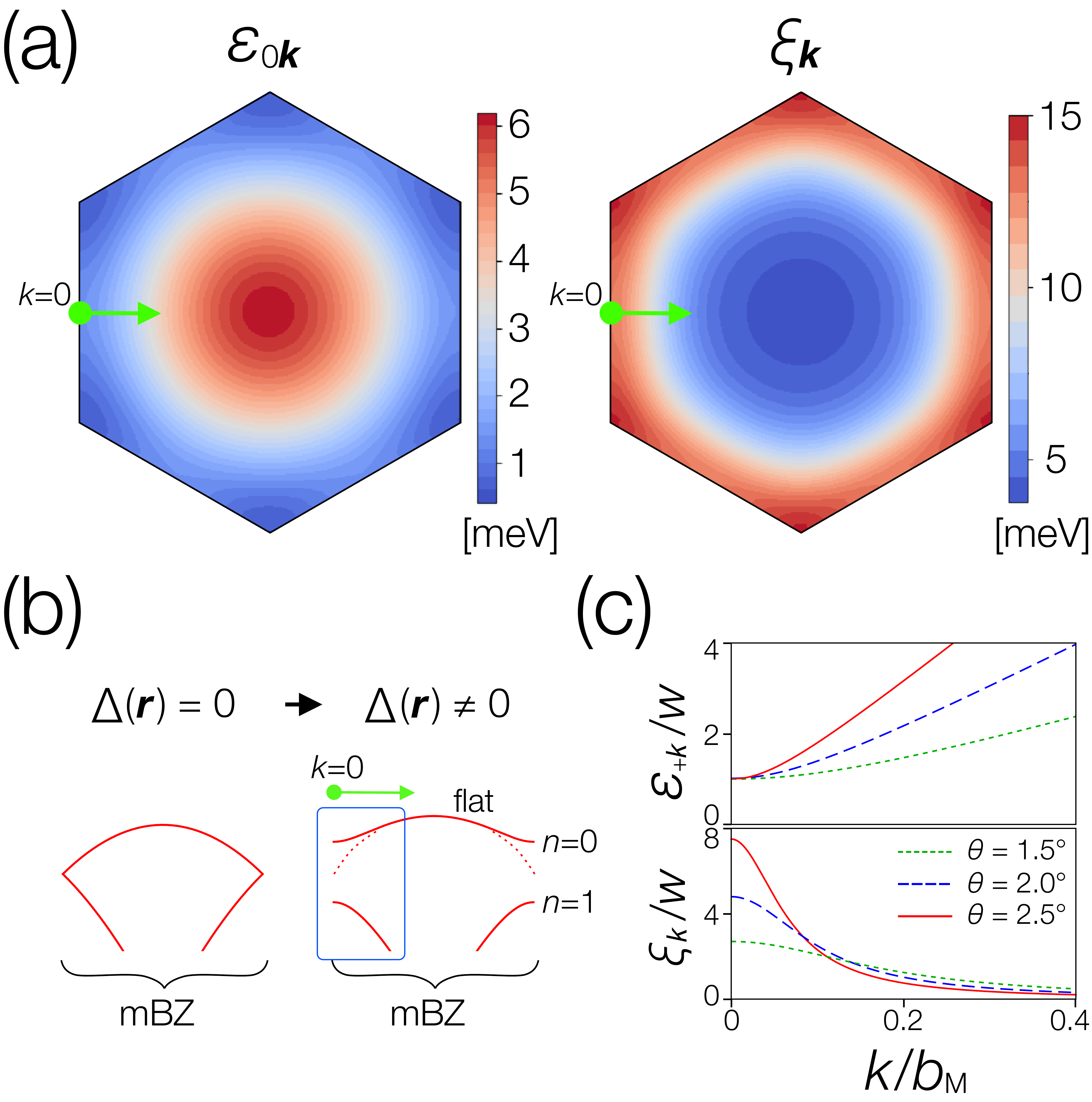}
    \caption{\label{figure3}(a) Typical moir\'e flat-band dispersion \(\varepsilon_{0\bm k}\) (left panel) and cavity renormalization \(\xi_{\bm k}\) in Eq.~\eqref{xi_k} (right panel) on the mBZ with \(\theta\!=\!1.7^\circ\) and \(\hbar\omega_{\bm 0}\!=\!10^2\) meV. (b) Schematic figure illustrating the formation of the moir\'e flat band. (c) Band dispersion \(\varepsilon_{+k}\) and cavity renormalization \(\xi_k\) (Eqs.~\eqref{2_band_ham_energy} and~\eqref{2_band_ham_xi_k}) of the two-band Hamiltonian~\eqref{2_band_ham} plotted against \(k/b_M\) with \(b_M = 4\pi/(\sqrt{3}a_M)\), where \(k\!=\!0\) corresponds to the edge of the mBZ as indicated in (a) and (b).}
\end{figure}

To reveal generic features of cavity moir\'e materials, in Fig.~\ref{figure3}(a) we show typical bare dispersion \(\varepsilon_{0\bm k}\) of the nearly flat band and the corresponding cavity renormalization \(\xi_{\bm k}\) on the mBZ. 
Interestingly, the dispersions of \(\varepsilon_{0\bm k}\) and \(\xi_{\bm k}\) are opposite each other, namely, \(\varepsilon_{0\bm k}\) (\(\xi_{\bm k}\)) takes the largest value at the center (edge) of the mBZ. Moreover, the contributions of \(\xi_{\bm k}\) are tightly localized around the edge of the mBZ, which can be translated to the emergence of long-range electron hoppings in real-space basis.

To understand these key features, we first note that the coefficient, \(\hbar\omega_{\bm 0}/(\hbar\omega_{\bm 0}+\varepsilon_{0\bm k}-\varepsilon_{n\bm k})\), in the renormalization \(\xi_{\bm k}\) in Eq.~\eqref{xi_k} monotonically decreases with respect to the band index \(n\), and the cavity dressing \(\xi_{\bm k}\) is mainly attributed to electronic bands with small \(n\). Therefore, the behavior of \(\varepsilon_{0\bm k}\) and \(\xi_{\bm k}\) can be qualitatively understood by analyzing a simple two-band model for moir\'e electronic bands with \(n=0,1\). In the TMD bilayer, the original monolayer band around VBM is first folded into the mBZ (left panel in Fig.~\ref{figure3}(b)) in accordance with the change of the lattice constant from \(a_0\) to \(a_M\).  The degeneracy at the edge of the mBZ is then lifted by the moir\'e potential \(\Delta(\bm r)\), leading to the nearly flat moir\'e band (right panel in Fig.~\ref{figure3}(b)). We can thus describe the energy bands near the mBZ edge by the following two-band Hamiltonian:
\begin{align}
	\mathcal{H}(k) = \hbar v_Fk \hat\sigma_z + w\hat\sigma_x,\label{2_band_ham}
\end{align}
where \(\hat \sigma_i\) is the Pauli matrix, \(k\) is the wavevector measured from the edge of the mBZ, \(v_F\) is the Fermi velocity at the mBZ edge (cf. left panel in Fig.~\ref{figure3}(b)), and \(w\!=\!|w_j|\) is the depth of the moir\'e potential in Eq.~\eqref{moi_potential}.
Using the two-band model~\eqref{2_band_ham}, the energy dispersions \(\varepsilon_{\pm k}\) and the cavity renormalization \(\xi_k\) can be obtained as
\begin{align}
	\varepsilon_{\pm k} & = \pm w \sqrt{1 + \left(\frac{\hbar v_Fk}{w}\right)^2},\label{2_band_ham_energy}                  \\
	\xi_k               & = \frac{m^* v_F^2}{1 + \left(\frac{\hbar v_Fk}{w}\right)^2} + {\rm const.}\label{2_band_ham_xi_k}
\end{align}
As shown in Fig.~\ref{figure3}(c), \(\varepsilon_{+k}\) (\(\xi_k\)) takes the smallest (largest) value at the edge of the mBZ, which correctly reproduces the qualitative features in Fig.~\ref{figure3}(a). Also,  \(\xi_k\) in Eq.~\eqref{2_band_ham_xi_k} is localized around \(k\!=\!0\) with a width \(\Delta k\!\sim\!w/(\hbar v_F)\propto \theta^{-1}\) (cf. Fig.~\ref{figure3}(c)). In real space, this leads to a long-distance hopping whose range is proportional to $\theta$. In general, such long-range contribution can generate strong magnetic frustration and is expected to qualitatively affect the ground-state properties as discussed below.

\section{Tight-binding description and the spin model\label{sec:spin-model}}

Using the Wannier basis, we can rewrite the effective Hamiltonian~\eqref{cav_moi_ham3} by the following Hubbard model on the moir\'e triangular lattice:
\begin{align}
	\hat H_{\rm eff} & = \sum_{ij,\sigma} t_{ij} \hat c_{i\sigma}^\dagger\hat c_{j\sigma} + U \sum_i \hat n_{i\uparrow}\hat n_{i\downarrow},\label{hubbard} \\
	t_{ij}           & = \int_{\rm mBZ} \frac{v_M d^2\bm k}{(2\pi)^2} (\varepsilon_{0\bm k} + \eta^2 \xi_{\bm k}) e^{i\bm k(\bm R_i-\bm R_j)}               \\
	                 & \equiv \tilde{\varepsilon}_{ij} + \eta^2 \tilde{\xi}_{ij},\label{hopping_amp}
\end{align}
where  \(\hat c_{i\sigma}^{(\dagger)}\) is the annihilation (creation) operator of the Wannier orbital localized at \(\bm R_{i}\), \(v_M\) is the volume of the moir\'e supercell, and we simplify the Coulomb repulsion by the on-site interaction \(U\sum_i\hat n_{i\uparrow}\hat n_{i\downarrow}\) \cite{wu2018Hubbard}. Let us represent the hoppings \(t_{ij}\) to the \(n\)-th nearest neighbors (NNs) by \(t_n\) and neglect the amplitudes \(t_n\) with \(n\!\geq\!4\), which are sufficiently small when \(\theta \lesssim 2.5^\circ\). Since the parameters satisfy  \(t_{n} \ll U\)~\cite{wu2018Hubbard}, at half-filling, we can further simplify the Hubbard model~\eqref{hubbard} to the triangular AFM Heisenberg model \cite{macdonald1988FractU}:
\begin{align}
	\hat H_{\rm AFM} & = \sum_{1\text{st NN}} J_1 \hat{\bm s}_i\!\cdot\!\hat {\bm s}_j + \!\sum_{2\text{nd NN}} J_2 \hat{\bm s}_i\!\cdot\!\hat {\bm s}_j +\!\sum_{3\text{rd NN}} J_3 \hat{\bm s}_i\!\cdot\!\hat {\bm s}_j,\label{heisenberg}
\end{align}
where \(\hat{\bm s}_i\) is the spin-\(1/2\) operator at the \(i\)-th site, and \(J_n \equiv 4t_{n}^2/U\). This frustrated spin model has been studied in, e.g., Refs.~\cite{zhu2015Spin,gong2019Chirala,hu2019Dirac,szasz2020Chiral,drescher2022Dynamical}, and a nonmagnetic insulating phase, which is a candidate QSL phase, is found in the parameter region around \(J_2/J_1\!\sim\! 0.3\) and \(J_3/J_1\!\sim\!0.15\)~\cite{gong2019Chirala}.

\begin{figure}[t]
	\includegraphics[width = 8.6cm]{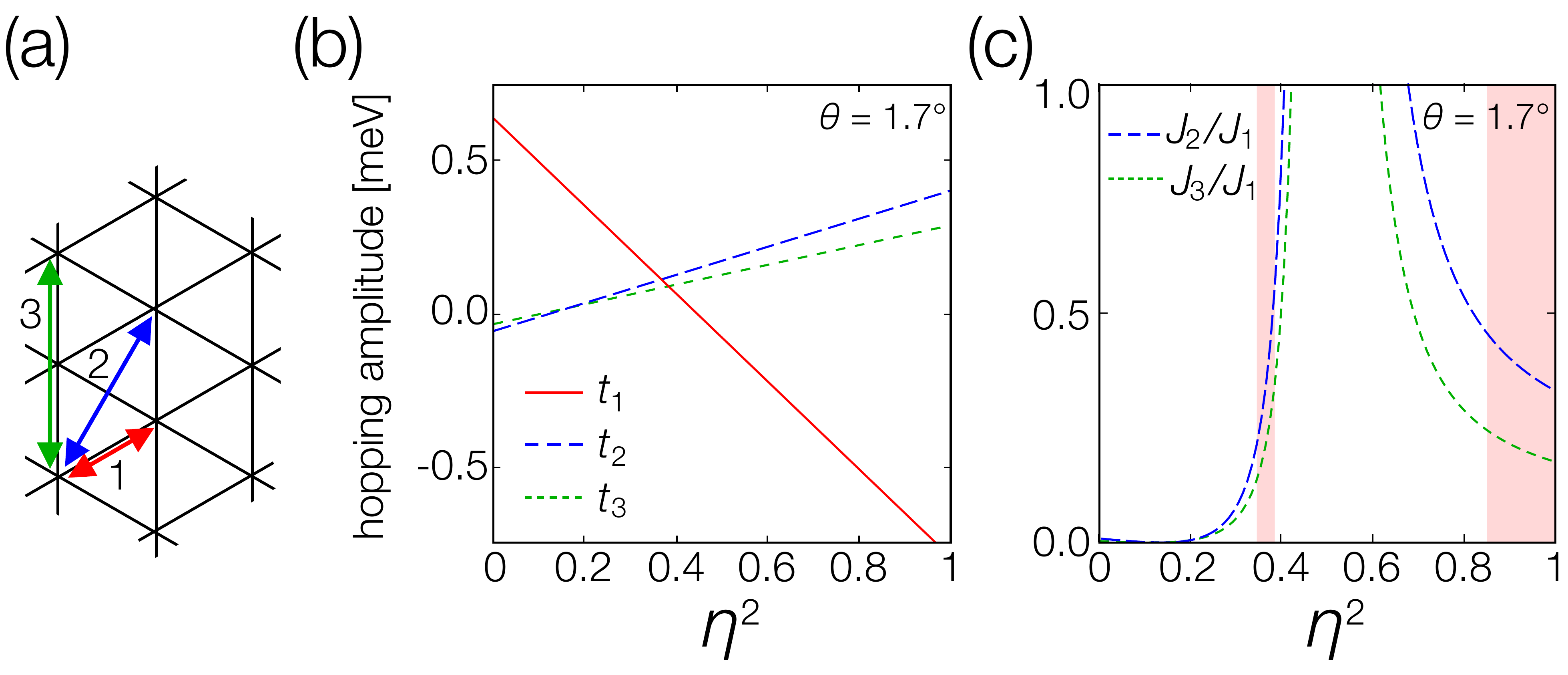}
	\caption{\label{figure4}(a) First-, second-, and third-order nearest neighbors (NNs) on the moir\'e triangular lattice. (b) Renormalized hopping amplitudes \(t_n\) to the \(n\)-th NNs and (c) corresponding spin-coupling ratios \(J_i/J_1\) (\(i = 2,3\)) in the triangular AFM Heisenberg model~\eqref{heisenberg} plotted against the light-matter coupling strength at \(\theta\!=\!1.7^\circ\). The red shaded areas in (c) indicate the intermediate parameter regions where the ground state is expected to be a quantum spin liquid.}
\end{figure}

One of our main findings is the possibility of controlling such spin frustration by the cavity confinement. Specifically, one can tune the light-matter coupling strength \(\eta\) to enhance the hopping amplitudes \(t_n\) to distant sites (\(n\!=\!2,3\)) while suppressing the NN hopping $t_1$ (Fig.~\ref{figure4}(b)). Physically, this tunability originates from the opposite dispersions in \(\varepsilon_{0\bm k}\) and \(\xi_{\bm k}\) described above (Fig.~\ref{figure3}(a)), which correspond to the opposite signs of their Fourier transforms, \(\tilde\varepsilon_{n}\) and \(\tilde\xi_{n}\),  that are related to the renormalized hopping via \(t_n = \tilde\varepsilon_{n} + \eta^2 \tilde\xi_{n}\).
We show the corresponding $\eta$-dependence of \(J_2/J_1\) and \(J_3/J_1\) in Fig.~\ref{figure4}(c), where one can realize particularly large spin-coupling ratios \(J_n/J_1\) when the NN coupling \(J_1\) becomes vanishingly small. Comparing with the numerical results in the triangular AFM Heisenberg model~\cite{gong2019Chirala}, we determine the ground-state phase diagram of the present cavity moir\'e system as in Fig.~\ref{figure1}(b). The key finding is that the high tunability of $\eta$ and $\theta$ should allow for controlling various correlated phases including a candidate QSL phase.

\section{Discussions and conclusions\label{sec:Discussion}}
We recall that a central feature of cavity moir\'e materials is the renormalization of single-electron energies in flat bands, which is induced by the virtual interband transitions originating from the nontrivial quantum geometry of moir\'e bands (cf. Eqs.~\eqref{xi_k} and \eqref{G_mn}).
While vacuum fluctuations can also induce the Amperean electron-electron interaction~\cite{schlawin2019CavityMediated}, we note that this contribution is negligibly small compared to the Coulomb interaction due to the flatness of moir\'e bands as detailed in Appendix~\ref{sec:app:derivation}.
In contrast, a usual monolayer has dispersive bands and band gap is typically much larger than a resonance frequency of the \textit{h}-BN cavity considered here. Thus, if monolayer is confined in the vdW slabs, the Amperean paring term could be also relevant.

In summary, we developed a theory of moir\'e materials strongly coupled to quantized electromagnetic fields inside a cavity. We showed that the major effect of the cavity confinement is the renormalization of the single-electron energies in flat bands. Physically, this originates from the nontrivial quantum geometry of the moir\'e bands, which leads to virtual interband electronic transitions induced by electromagnetic vacuum fluctuations. The resulting long-range electronic hoppings allow for tuning the spin frustration in the low-energy effective model, thus suggesting the possibility of controlling correlated phases with quantum light-matter interaction. We analyzed the concrete setup consisting of TMD heterobilayer WSe\(_2\)/MoSe\(_2\) confined in the \textit{h}-BN cavity, and revealed that various phases, including a putative QSL state, can be realized by varying the light-matter coupling strength \(\eta\) (Fig.~\ref{figure1}(b)).

Ultimately, a further nonperturbative analysis should be necessary to fully assess the validity of our predictions in the relevant interaction strength regime in Fig.~\ref{figure1}(b) around \(\eta^2 \sim 0.4\), while to our knowledge such a rigorous nonperturbative framework for studying cavity moir\'e materials with the continuum of quantized electromagnetic modes is currently lacking. Nevertheless, we argue that the effect we predict, the enhancement of magnetic frustration, is expected to be appreciable even in this regime. The reason for this is that the enhancement is based on a simple and general mechanism, namely, the emergence of long-range spin interactions mediated by the vacuum fluctuations of cavity fields; at this stage, we do not expect any particular reason for breakdown of this mechanism in nonperturbative regions. In addition, it is also important to note that the essence of controlling magnetic phases of moir\'e materials is the band renormalization \(\xi_{\bm k}\) in the effective Hamiltonian~\eqref{cav_moi_ham3} which is larger than the original bare dispersion \(\varepsilon_{\bm k}\) due to the band flatness. Therefore, the effect we predict becomes more appreciable in moir\'e materials with flatter electronic bands, and the transition might already take place in perturbative regimes. All in all, our findings would be of experimental relevance in view of recent developments in moir\'e materials and cavity QED, which demonstrates ultrasmall mode volumes of hyperbolic polaritons in nanostructured materials \cite{CJD14,SD14,GA18,SD19,MEY22,SHH22}.

Several open questions remain for future studies. In the present perturbation theory, we only retain the leading terms of \({\rm O}(\eta^2)\), while the higher-order corrections might be important especially when \(\eta ={\rm O}(1)\). It would be interesting to analyze such a challenging regime on the basis of a nonperturbative approach allowing for nonvanishing electron-photon entanglement \cite{WY21,YA21}.
It is also worthwhile to recall that the cavity-mediated hoppings \(\eta^2\tilde\xi_{ij}\) in Eq.~\eqref{cav_moi_ham3} become more long-ranged as  twist angle \(\theta\) is increased. It merits further study to examine these large-\(\theta\) regimes of cavity moir\'e materials beyond the parameter region considered here \cite{LM20,ZL20,TS22}; there, higher-order hoppings $t_n$ should be increasingly important and an effective Hamiltonian may exhibit stronger magnetic frustration in which the ground-state phase diagram could be enriched. We hope that our work simulates further studies in these directions.

\begin{acknowledgments}
	We are grateful to Eugene Demler, Shunsuke Furukawa, Atac Imamoglu, Jun Mochida, Yao Wang, and Kenji Yasuda for fruitful discussions.
    K. M. acknowledges support from the Japan Society for the Promotion of Science (JSPS) through Grant No.~JP24KJ0898. Y.A. acknowledges support from the Japan Society for the Promotion of Science (JSPS) through Grant No. JP19K23424 and from JST FOREST Program (Grant Number JPMJFR222U, Japan) and JST CREST (Grant Number JPMJCR23I2, Japan). \\
\end{acknowledgments}

\appendix

\centerline{{\bf APPENDIX}}

\section{Derivation of the effective Hamiltonian of the flat-band electrons\label{sec:app:derivation}}

\begin{figure}[b]
	\includegraphics[width = 8.6cm]{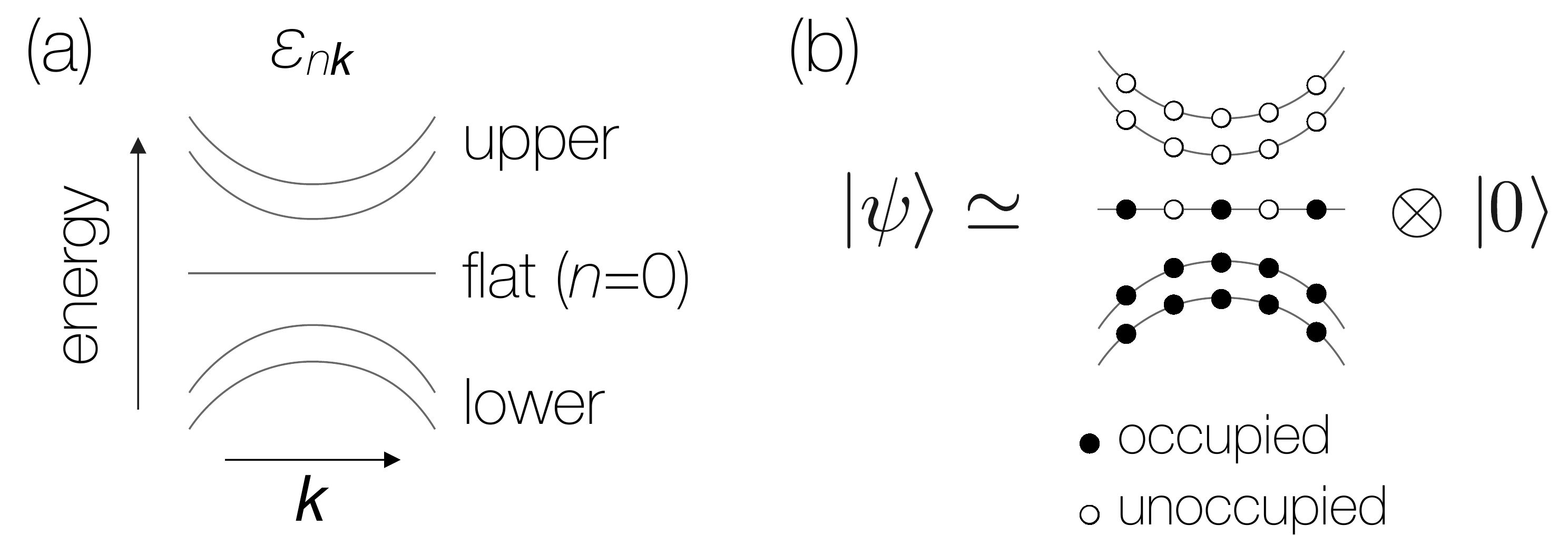}
	\caption{\label{sm_figure1}(a) Schematic band dispersions of a moir\'e Hamiltonian \(\hat H_0 = -\hat{\bm p}^2/2m^* + \Delta(\bm r)\), where the flat band is located between the other electronic bands. (b) Approximate low-energy states \(|\psi\rangle \simeq |\psi_{\rm e}\rangle\otimes|0\rangle\) consisting of the product between the electromagnetic vacuum \(|0\rangle\) and an electronic state \(|\psi_{\rm e}\rangle\) with the partially filled flat band and occupied (unoccupied) lower (upper) bands.}
\end{figure}

We here provide the detailed derivation of the effective Hamiltonian of flat-band electrons in Eq.~\eqref{cav_moi_ham3} in the main text. Specifically, we start from the cavity moir\'e Hamiltonian in Eq.~\eqref{cav_moi_ham1} in the main text.
For the sake of generality, we consider the band structure shown in Fig.~\ref{sm_figure1}(a) as the dispersions \(\varepsilon_{n\bm k}\) of the bare moir\'e Hamiltonian \(\hat H_0\!\equiv\!-\hat{\bm p}^2/2m^* + \Delta(\bm r)\), where the flat band (labelled by $n=0$) is located between the other electronic bands ($n\neq 0$). As described in the main text, the low-energy states of cavity moir\'e Hamiltonian~\eqref{cav_moi_ham1} can be approximated by a product state \(|\psi_{\rm e}\rangle\otimes|0\rangle\), where \(|0\rangle\) is the electromagnetic vacuum, and \(|\psi_{\rm e}\rangle\) is an electronic state with the partially-filled flat band and occupied (unoccupied) lower (upper) bands (cf. Fig.~\ref{sm_figure1}(b)). Accordingly, we define the projection operator \(\hat P\) onto the manifold spanned by these states and obtain the effective Hamiltonian in this subspace by employing the perturbation theory with respect to the (dimensionless) light-matter coupling strength \(\eta = \sqrt{\sum_{\bm q}g_{\bm q}^2/\omega_{\bm q}^2}\).

\begin{widetext}
To proceed, we decompose the cavity moir\'e Hamiltonian~\eqref{cav_moi_ham3} as
\begin{align}
	\hat H =           & \hat H_0 + \sum_{\bm q}\hbar\omega_{\bm q}'\hat a_{\bm q}^\dagger\hat a_{\bm q} + \hat H_I^{(p)} + \hat H_I^{(q)} + \hat U,\label{sm_cav_moi_ham2}\\
	\omega_{\bm q}' =  & \omega_{\bm q}\left(1 - \frac{N_ee^2A_{\bm q}^2}{m^*\hbar\omega_{\bm q}}\right) = \omega_{\bm q}\left(1 - \frac{N_eg_{\bm q}^2}{\omega_{\bm q}^2}\right),\label{sm_cavfreq_renorm}\\
	\hat H_{I}^{(p)} = & - \sum_{\bm q} \frac{eA_{\bm q}}{m^*} \left(\sum_{\sigma}\int d^2{\bm r}\ \hat\psi_{\sigma\bm r}^\dagger \frac{(\hat{\bm p}\!\cdot\!\bm e_{\bm q})e^{i\bm q\bm r} + e^{i\bm q\bm r}(\hat{\bm p}\!\cdot\!\bm e_{\bm q})}{2} \hat\psi_{\sigma\bm r} \right)\hat a_{\bm q} + {\rm H.c.},  \\
	\hat H_{I}^{(q)} = & \sum_{\bm q_1,\bm q_2} \sum_{\sigma} \int d^2{\bm r}\ \hat\psi_{\sigma\bm r}^\dagger \frac{e^2A_{\bm q_1}A_{\bm q_2}}{2m^*} \bm e_{\bm q_1}\!\cdot\! \bm e_{\bm q_2} (\hat a_{\bm q_1}\hat a_{\bm q_2}e^{i(\bm q_1+\bm q_2)\bm r} + {\rm H.c.})\ \hat\psi_{\sigma\bm r} \nonumber      \\
	                   & + \sum_{\bm q_1\neq\bm q_2} \sum_{\sigma} \int d^2{\bm r}\ \hat\psi_{\sigma\bm r}^\dagger \frac{e^2A_{\bm q_1}A_{\bm q_2}}{2m^*} \bm e_{\bm q_1}\!\cdot\! \bm e_{\bm q_2} (\hat a_{\bm q_1}^\dagger\hat a_{\bm q_2}e^{-i(\bm q_1-\bm q_2)\bm r} + {\rm H.c.})\ \hat\psi_{\sigma\bm r},
\end{align}
\end{widetext}
where \(\hat H_I^{(p)}\) is the paramagnetic term, and the $A^2$ diamagnetic term leads to the renormalized cavity frequency \(\omega_{\bm q}'\) and the interaction term \(\hat H_{I}^{(q)}\). In Eq.~\eqref{sm_cavfreq_renorm}, we define the total number of electrons by \(N_e = \sum_{\sigma}\int d^2{\bm r}\hat\psi_{\sigma\bm r}^\dagger\hat\psi_{\sigma\bm r}\), which is an order of \(V/v_M\) with the system size \(V\) and the volume of the moir\'e supercell \(v_M\). Since the coupling strength \(g_{\bm q}\) is proportional to \(\sqrt{1/V}\), the frequency renormalization in Eq.~\eqref{sm_cavfreq_renorm} is \({\rm O}(V^0\eta^2)\). Below, we treat the first two terms in the right hand side of Eq.~\eqref{sm_cav_moi_ham2} as the unperturbed Hamiltonian and incorporate the effect of \(\hat H_{I}^{(p)} + \hat H_{I}^{(q)}\) by the perturbation theory.

\begin{figure*}[t]
	\includegraphics[width = 16cm]{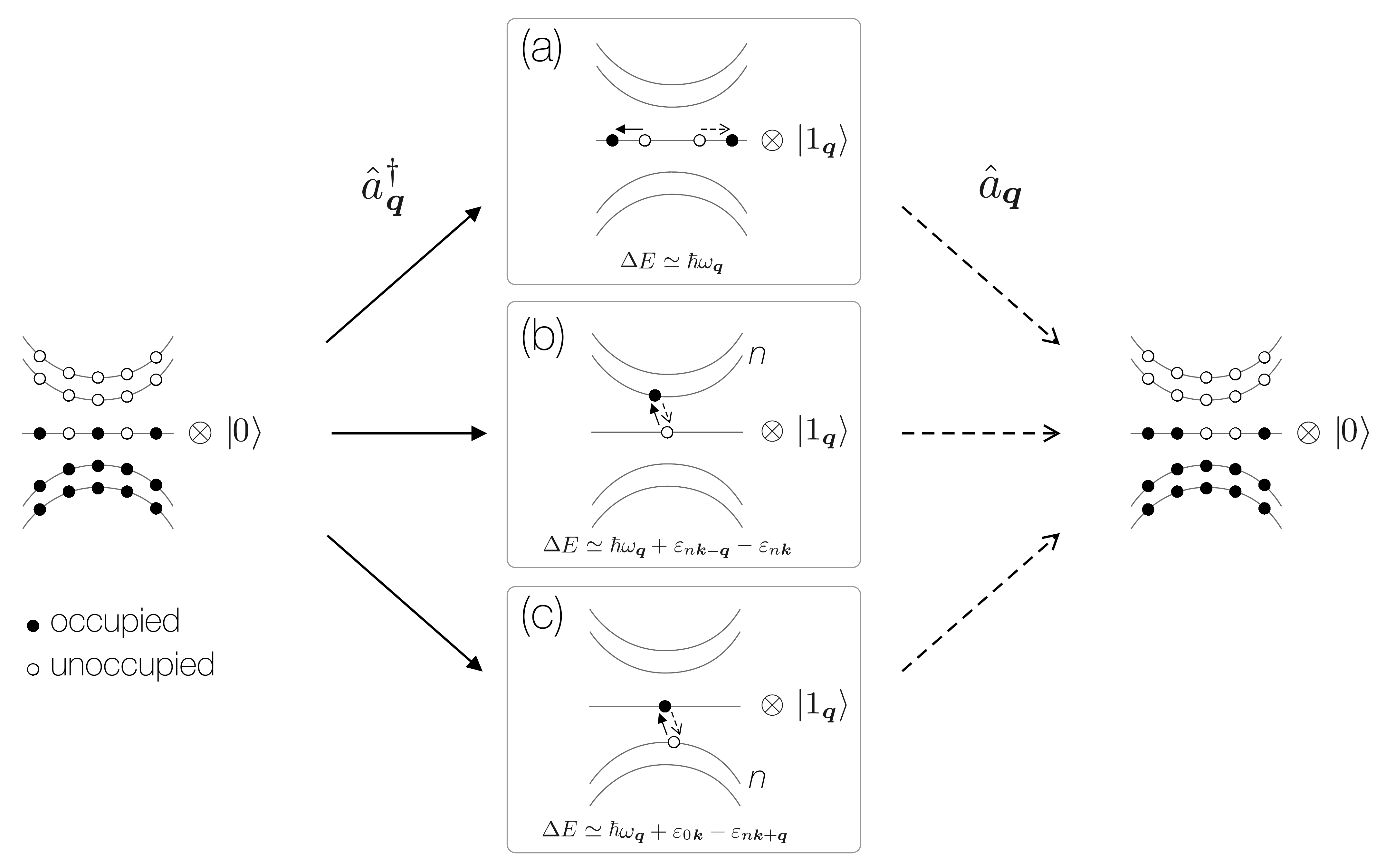}
	\caption{\label{sm_figure2} Virtual processes in the second-order perturbation of the paramagnetic term \(\hat H_I^{(p)}\), where a cavity mode is excited by \(\hat a_{\bm q}^\dagger\) with (a) intraband transitions within the flat band, (b) interband transitions from the flat band to an upper band, and (c) interband transitions from a lower band to the flat band. We approximate excitation energies in each process \(\Delta E\) as shown in (a)-(c).}
\end{figure*}

Since \(\hat P\hat H_I^{(p)}\hat P = \hat P\hat H_I^{(q)}\hat P = 0\), the leading terms come from the second-order perturbation of the paramagnetic term \(\hat H_I^{(p)}\), which can be expressed in the Bloch basis as
\begin{widetext}
\begin{align}
	\hat H_I^{(p)} & = - \sum_{\bm k\sigma}\sum_{mn}\sum_{\bm q} \frac{eA_{\bm q}}{m^*}\bm e_{\bm q}\!\cdot\! \langle \psi_{m\bm k+\bm q\sigma}|\frac{\hat{\bm p}e^{i\bm q\bm r} + e^{i\bm q\bm r}\hat{\bm p}}{2}|\psi_{n\bm k\sigma}\rangle \hat c_{m\bm k+\bm q\sigma}^\dagger\hat c_{n\bm k\sigma}\hat a_{\bm q} + {\rm H.c.} \\
    & = - \sum_{\bm k\sigma}\sum_{mn}\sum_{\bm q} \frac{eA_{\bm q}}{m^*}\bm e_{\bm q}\!\cdot\! \langle u_{m\bm k+\bm q}|\left(\hat{\bm p} + \hbar\left(\bm k + \frac{\bm q}{2}\right) \right)|u_{n\bm k}\rangle \hat c_{m\bm k+\bm q\sigma}^\dagger\hat c_{n\bm k\sigma}\hat a_{\bm q} + {\rm H.c.}               \\
    & = - \sum_{\bm k\sigma}\sum_{mn}\sum_{\bm q} \frac{eA_{\bm q}}{\hbar}(\bm G_{mn}^{\bm k,\bm q}\!\cdot\!\bm e_{\bm q})\ \hat c_{m\bm k+\bm q\sigma}^\dagger\hat c_{n\bm k\sigma}\hat a_{\bm q} + {\rm H.c.}
\end{align}
\end{widetext}
Here, we introduce the annihilation (creation) operator \(\hat c_{n\bm k\sigma}^{(\dagger)}\) corresponding to the Bloch state \(|\psi_{n\bm k\sigma}\rangle\) with the spin index $\sigma$ and also define the multiband coefficients \(\bm G_{mn}^{\bm k,\bm q} \equiv \langle u_{m\bm k+\bm q}|\nabla_{\bm k}\hat H_{\bm k+\bm q/2}|u_{n\bm k}\rangle\), where \(|u_{n\bm k}\rangle\) is related to \(|\psi_{n\bm k\sigma}\rangle\) via \(|\psi_{n\bm k\sigma}\rangle\!=\!e^{i\bm k\hat{\bm r}}|u_{n\bm k}\rangle\!\otimes\!|\sigma\rangle\) with spin state \(|\sigma\rangle\). Figures~\ref{sm_figure2}(a)-(c) show the virtual processes relevant to the second-order perturbation of \(\hat H_{I}^{(p)}\), each of which corresponds to the virtual electronic transitions (a) within the flat band, (b) to the upper bands, and (c) from the lower bands, respectively. Approximating the excitation energy \(\Delta E\) in each process as shown in Fig.~\ref{sm_figure2}, we get the following effective Hamiltonian of flat-band electrons:
\begin{widetext}
\begin{align}
	\hat H_{\rm eff} =   & \sum_{\bm k\sigma} \varepsilon_{0\bm k} \hat c_{\bm k\sigma}^\dagger \hat c_{\bm k\sigma} + \hat H_{\rm ff} + \hat H_{\rm fu} + \hat H_{\rm fl} + \hat U,   \label{sm_H_eff}                                                                                                                                                                                            \\
	\hat H_{\rm ff} =    & - \sum_{\bm k,\bm k'}\sum_{\sigma,\sigma'} \sum_{\bm q} \frac{m^*}{\hbar^2} \frac{g_{\bm q}^2}{{\omega'_{\bm q}}^2} (\bm G_{00}^{\bm k,\bm q}\!\cdot\!\bm e_{\bm q})(\bm G_{00}^{\bm k',\bm q*}\!\cdot\!\bm e_{\bm q}) \hat c_{\bm k+\bm q\sigma}^\dagger\hat c^\dagger_{\bm k'-\bm q\sigma'}\hat c_{\bm k'\sigma'}\hat c_{\bm k\sigma}\nonumber        \\
	                     & -\sum_{\bm k\sigma} \frac{m^*}{\hbar^2} \left( \sum_{\bm q} \frac{g_{\bm q}^2}{{\omega'_{\bm q}}^2} |\bm G_{00}^{\bm k,\bm q}\!\cdot\!\bm e_{\bm q}|^2 \right) \hat c_{\bm k\sigma}^\dagger\hat c_{\bm k\sigma},                                                 \label{sm_H_ff}                                                                        \\
	\hat H_{\rm fu}    = & -\sum_{\bm k\sigma} \frac{m^*}{\hbar^2} \left( \sum_{\bm q}\sum_{n}^{\rm upper} \frac{g_{\bm q}^2}{{\omega'_{\bm q}}^2} \frac{\hbar\omega_{\bm q}'}{\hbar\omega_{\bm q}' + \varepsilon_{n\bm k+\bm q}-\varepsilon_{0\bm k}} |\bm G_{n0}^{\bm k,\bm q}\!\cdot\!\bm e_{\bm q}|^2 \right) \hat c_{\bm k\sigma}^\dagger\hat c_{\bm k\sigma},\label{sm_H_fu} \\
	\hat H_{\rm fl}    = & \sum_{\bm k\sigma} \frac{m^*}{\hbar^2} \left( \sum_{\bm q}\sum_{n}^{\rm lower} \frac{g_{\bm q}^2}{{\omega'_{\bm q}}^2} \frac{\hbar\omega_{\bm q}'}{\hbar\omega_{\bm q}' + \varepsilon_{0\bm k}-\varepsilon_{n\bm k+\bm q}} |\bm G_{0n}^{\bm k,\bm q}\!\cdot\!\bm e_{\bm q}|^2 \right) \hat c_{\bm k\sigma}^\dagger\hat c_{\bm k\sigma},\label{sm_H_fl}
\end{align}
\end{widetext}
where \(\hat c_{\bm k\sigma}^{(\dagger)} = \hat c_{0\bm k\sigma}^{(\dagger)}\) is the annihilation (creation) operator of the flat-band electrons, and \(\sum_{n}^{\rm upper(lower)}\) denotes the summation over the upper (lower) bands. In deriving Eqs.~\eqref{sm_H_ff}-\eqref{sm_H_fl}, we assume the relations \(g_{\bm q}=g_{-\bm q}\) and \(\omega_{\bm q}' = \omega'_{-\bm q}\) that hold true for uniaxial cavities, and use the following equalities:
\begin{widetext}
\begin{align}
	\hat P(\hat c_{0\bm k\sigma}^\dagger \hat c_{n\bm k+\bm q\sigma}\hat c_{n'\bm k'+\bm q\sigma'}^\dagger \hat c_{0\bm k'\sigma'})\hat P = \delta_{nn'}\delta_{\bm k\bm k'}\delta_{\sigma\sigma'} \hat c_{0\bm k\sigma}^\dagger\hat c_{0\bm k\sigma}\ (n,n':\text{upper bands}), \\
	\hat P(\hat c_{n\bm k+\bm q\sigma}^\dagger \hat c_{0\bm k\sigma}\hat c_{0\bm k'\sigma'}^\dagger \hat c_{n'\bm k'+\bm q\sigma'})\hat P = \delta_{nn'}\delta_{\bm k\bm k'}\delta_{\sigma\sigma'} \hat c_{0\bm k\sigma}\hat c_{0\bm k\sigma}^\dagger\ (n,n':\text{lower bands}).
\end{align}
\end{widetext}
We note that the first term in the right hand side of Eq.~\eqref{sm_H_ff} is the Amperean electron-electron interaction mediated by cavity electromagnetic fields~\cite{schlawin2019CavityMediated}. Since the coefficients \((\bm G_{00}^{\bm k,\bm q}\!\cdot\!\bm e_{\bm q})(\bm G_{00}^{\bm k',\bm q*}\!\cdot\!\bm e_{\bm q})\approx(\nabla_{\bm k}\varepsilon_{0\bm k}\!\cdot\!\bm e_{\bm q})(\nabla_{\bm k}\varepsilon_{0\bm k'}\!\cdot\!\bm e_{\bm q})\) in this term almost vanish due to the band flatness, the Amperean interaction is negligibly small compared to the Coulomb interaction \(\hat U\). Retaining up to \({\rm O}(\eta^2)\)-terms and taking the limit \(\bm q\to\bm 0\) in Eq.~\eqref{sm_H_eff} (which does not affect the results provided in the present work as detailed in Appendix~\ref{sec:app:approx}), we finally obtain the effective Hamiltonian of flat-band electrons as
\begin{widetext}
\begin{align}
	\hat H_{\rm eff} & = \sum_{\bm k\sigma} (\varepsilon_{0\bm k} + \eta^2\xi_{\bm k}) \hat c_{\bm k\sigma}^\dagger\hat c_{\bm k\sigma} + \hat U,\label{sm_cav_moi_ham3}                                                                                                                                                                                                \\
	\xi_{\bm k}      & = \frac{m^*}{2\hbar^2} \left( -|\bm G_{00}^{\bm k}| - \sum_{n}^{\rm upper} \frac{\hbar\omega_{\bm 0}}{\hbar\omega_{\bm 0} + \varepsilon_{n\bm k}-\varepsilon_{0\bm k}} |\bm G_{n0}^{\bm k}|^2 + \sum_{n}^{\rm lower} \frac{\hbar\omega_{\bm 0}}{\hbar\omega_{\bm 0} + \varepsilon_{0\bm k}-\varepsilon_{n\bm k}} |\bm G_{n0}^{\bm k}|^2 \right),
\end{align}
\end{widetext}
where we introduce \(\bm G_{mn}^{\bm k}\equiv \bm G_{mn}^{\bm k,\bm 0}\) and use the fact that \(\omega_{\bm q}'/\omega_{\bm q} = 1 + {\rm O}(\eta^2)\). Since there are no upper bands in the setup of the TMD heterobilayer considered in the main text, the effective Hamiltonian~\eqref{sm_cav_moi_ham3} simplifies to Eq.~\eqref{cav_moi_ham3} in the main text.
We note that the effective Hamiltonian derived above is qualitatively different from the one obtained by using the single-mode approximation. In the latter case, the single-mode coupling \(g/\omega = O(V^0)\) would be required to achieve the  ultrastrong coupling \(g^2/\omega^2 =  O(1)\), and thus, the renormalization of the cavity frequency due to \(A^2\) diamagnetic term should be \(O(V^1)\). Also, we note that possible modification of the Coulomb interaction \(U\), which is at most \(O(\eta^2)\), is not included since it does not affect the discussions in the present work as detailed in Appendix~\ref{sec:app:Coulomb}.

\begin{figure}[t]
	\includegraphics[width = 7cm]{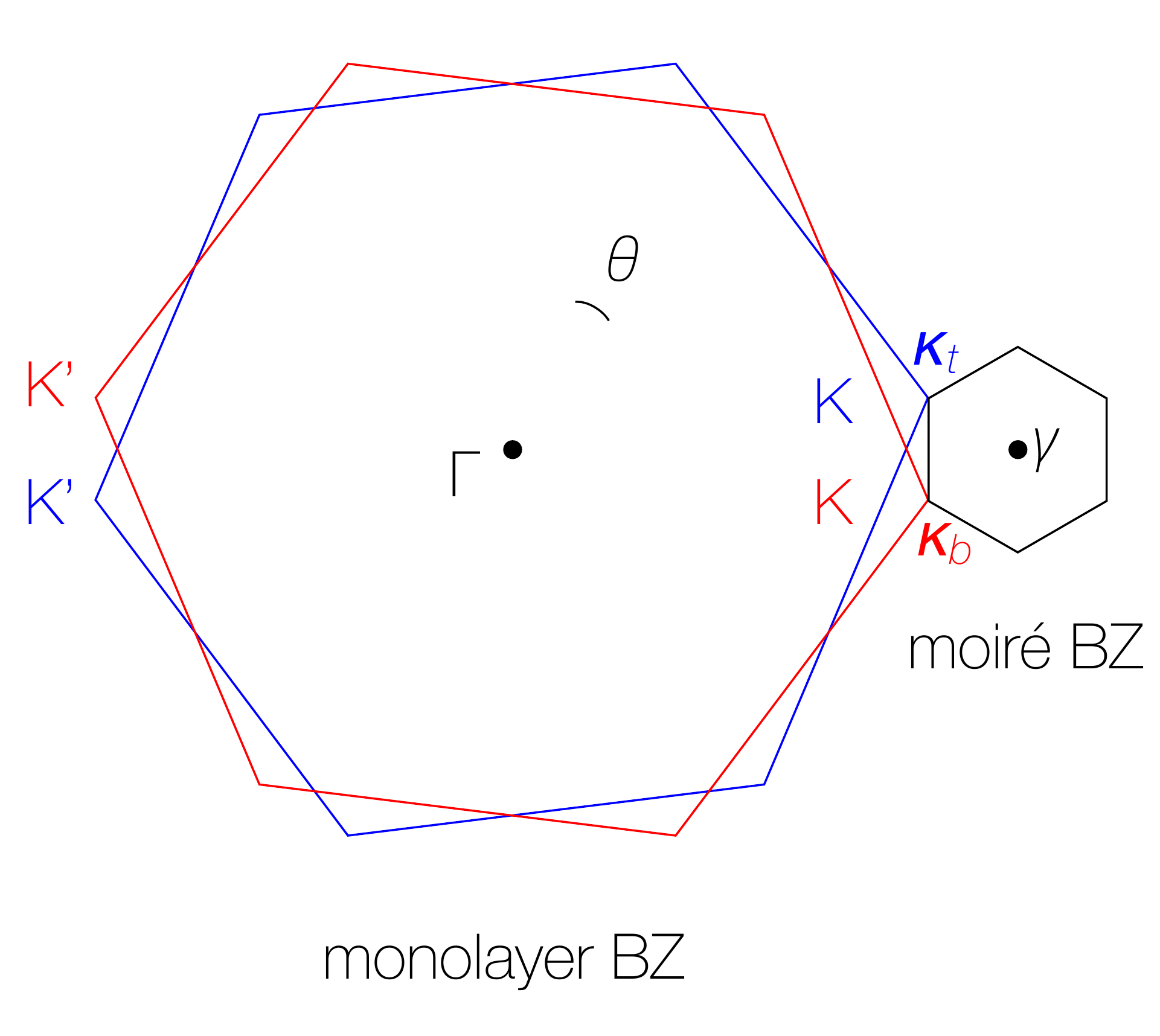}
	\caption{\label{sm_figure3}Moir\'e Brillouin Zone (mBZ) for spin-up (K-valley) electrons of TMD moir\'e homobilayer with twist angle \(\theta\). We also show the monolayer BZ of the top (blue) and bottom (red) TMD layers.}
\end{figure}

Making appropriate modifications in the moir\'e Hamiltonian~\eqref{moi_hamiltonian}, our procedure can also be applied to other moir\'e materials, such as a TMD moir\'e homobilayer or a twisted bilayer graphene. In a TMD moir\'e homobilayer, the moir\'e Hamiltonian of spin-up electrons (or equivalently, K-valley electrons in the TMD monolayers) is given as~\cite{carr2020Electronicstructure,WM2019-2},
\begin{align}
	\hat H_{0\uparrow} = \left(
	\begin{array}{cc}
			- \frac{(\hat{\bm p}-\hbar\bm \kappa_b)^2}{2m^*} &                                                  \\
			                                                 & - \frac{(\hat{\bm p}-\hbar\bm \kappa_t)^2}{2m^*}
		\end{array}
	\right) + \hat \Delta(\bm r),\label{sm_moi_ham_homo}
\end{align}
where the \(2\!\times\!2\)-matrix corresponds to the two layer degrees of freedom, and \(\hat\Delta(\bm r)\) is a \(2\!\times\!2\)-matrix-valued moir\'e potential. Here, \(\bm \kappa_{t(b)}\) denotes the distinct mBZ corner, which originates from the K-valley of the top (bottom) TMD monolayer (cf. Fig.~\ref{sm_figure3}). The moi\'e Hamiltonian of spin-down electrons \(\hat H_{0\downarrow}\) is given as the time-reversal (TR) pair of \(\hat H_{0\uparrow}\). We note that \(\hat H_{0\uparrow(\downarrow)}\) does not have TR-symmetry \((\hat{\bm p},\hat\Delta(\bm r))\to(-\hat{\bm p},\hat\Delta^*(\bm r))\), while the total moir\'e Hamiltonian \(\hat H_{0\uparrow} + \hat H_{0\downarrow}\) does \((\hat{\bm p},\hat\Delta(\bm r),\sigma)\to(-\hat{\bm p},\hat\Delta^*(\bm r),-\sigma)\). As a result, the hopping amplitudes \(t_{ij,\sigma}\) in the tight-binding description are in general complex-valued and satisfy \(t_{ij,-\sigma} = t_{ij,\sigma}^*\). Similarly, in a twisted bilayer graphene, the moir\'e Hamiltonian of spin-up and K-valley electrons is given as~\cite{carr2020Electronicstructure,bistritzer2011Moire},
\begin{align}
	\hat H_{0\uparrow\rm K} = \left(
	\begin{array}{cc}
			v_F(\hat{\bm p}-\hbar\bm \kappa_b)\!\cdot\!\hat{\bm \sigma} &                                                             \\
			                                                            & v_F(\hat{\bm p}-\hbar\bm \kappa_t)\!\cdot\!\hat{\bm \sigma} \\
		\end{array}
	\right) + \hat\Delta(\bm r),
\end{align}
where \(v_F\) is the Fermi velocity of monolayer graphene, \(\hat{\bm \sigma}=(\hat\sigma_x,\hat \sigma_y)^T\), and \(\hat\Delta(\bm r)\) is a \(4\!\times\!4\)-matrix-valued moir\'e potential corresponding to the layer and sublattice degrees of freedom.

\section{On the validity of the approximation made when deriving effective Hamiltonian of flat-band electrons\label{sec:app:approx}}

In the last step of the above derivation described in Appendix~\ref{sec:app:derivation}, we neglect the \(\bm q\)-dependences of $\bm G_{0n}^{\bm k,\bm q}$ and $\hbar\omega_{\bm q}/(\hbar\omega_{\bm q} + \varepsilon_{0\bm k} - \varepsilon_{n\bm k+\bm q})$ in Eqs.~\eqref{sm_H_ff}-\eqref{sm_H_fl} and approximate them as
\begin{widetext}
\begin{align}
    \hat H_{\rm fu} &= \sum_{\bm k\sigma} \frac{m^*}{\hbar^2} 
    \left( \sum_{\bm q}\sum_{n}^{\rm lower} \frac{g_{\bm q}^2}{\omega_{\bm q}^2} \frac{\hbar\omega_{\bm q}}{\hbar\omega_{\bm q} + \varepsilon_{0\bm k} - \varepsilon_{n\bm k+\bm q}}|\bm G_{0n}^{\bm k,\bm q}\cdot\bm e_{\bm q}|^2 \right) \hat c_{\bm k\sigma}^\dagger\hat c_{\bm k\sigma} + O(\eta^4)\\
    &\simeq \sum_{\bm k\sigma} \frac{m^*}{\hbar^2} 
    \left( \sum_{\bm q}\frac{g_{\bm q}^2}{\omega_{\bm q}^2} \sum_{n}^{\rm lower}  \frac{\hbar\omega_{\bm 0}}{\hbar\omega_{\bm 0} + \varepsilon_{0\bm k} - \varepsilon_{n\bm k}}|\bm G_{0n}^{\bm k,\bm 0}\cdot\bm e_{\bm q}|^2 \right) \hat c_{\bm k\sigma}^\dagger\hat c_{\bm k\sigma}\label{sm_approx}.
\end{align}
\end{widetext}
To demonstrate the validity of this approximation, we compare the main results in the main text obtained with or without making such approximation. Below, to include the effect of momentum dependence of the coupling strengths \(g_{\bm q}/\omega_{\bm q}\), we assume that \(g_{\bm q}^2/\omega_{\bm q}^2\) can be approximated by a simple Gaussian function as
\begin{align}
    \frac{g_{\bm q}^2}{\omega_{\bm q}^2} = \frac{\eta^2}{\mathcal{N}} \exp\left(-\frac{\bm q^2}{2q_c^2}\right),~\label{sm_coupling_strength}
\end{align}
where \(\mathcal N\) is determined by the normalization condition \(\sum_{\bm q} g_{\bm q}^2/\omega_{\bm q}^2 = \eta^2\). Also, we shall define the effective width \(Q\) of the coupling strength by the condition \(\sum_{|\bm q|<Q} g_{\bm q}^2/\omega_{\bm q}^2 = 0.95 \eta^2 \Leftrightarrow Q/q_c = \sqrt{-2\ln 0.05}\). 

\begin{figure*}
	\includegraphics[width = 14cm]{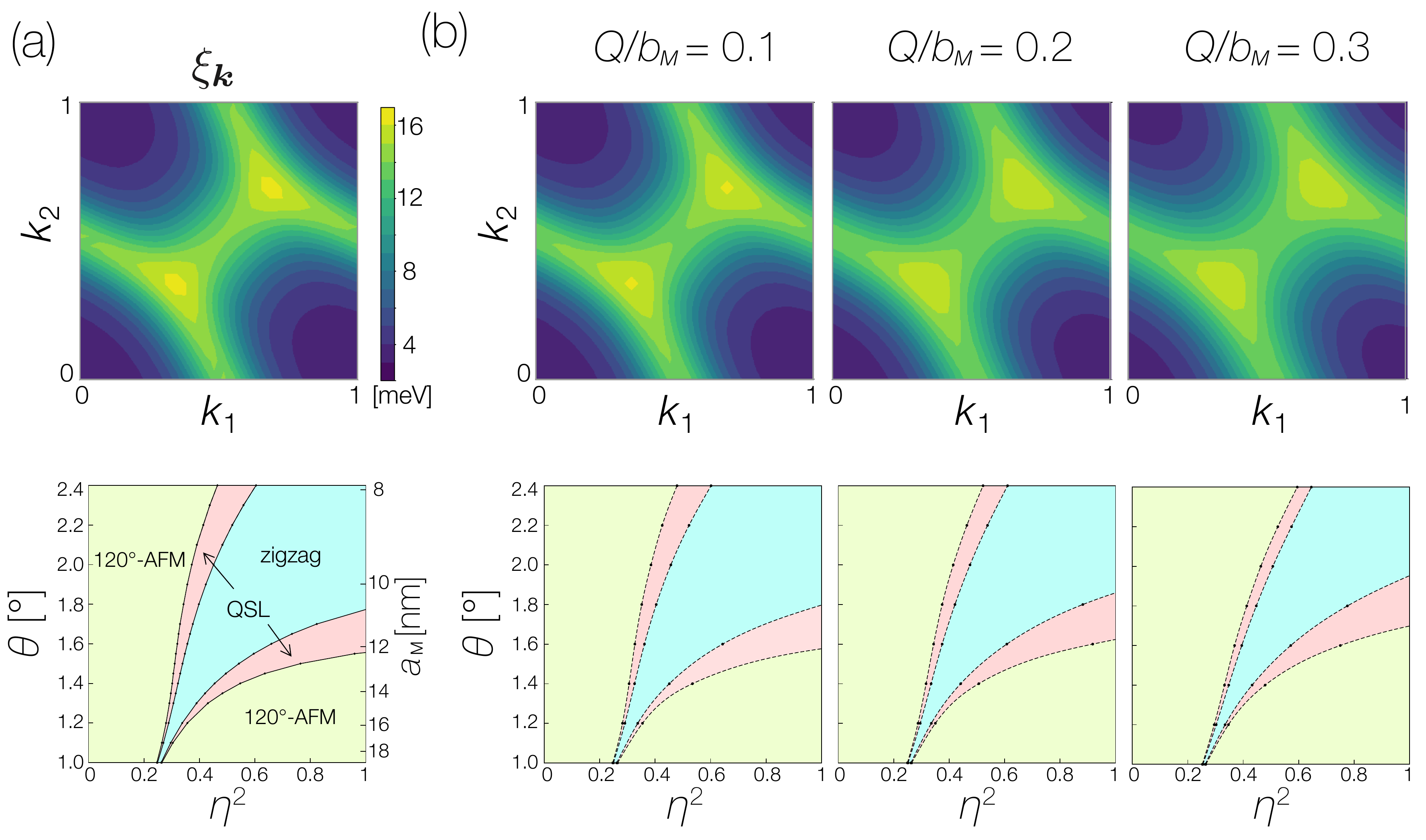}
	\caption{\label{sm_figure4}(a) Cavity dressing \(\xi_{\bm k}\) (top panel) and the ground-state phase diagram of the cavity-confined TMD heterobilayer (bottom panel) obtained with the approximation~\eqref{sm_approx}. We define \(k_1\) and \(k_2\) via \(\bm k = k_1\bm b_1 + k_2\bm b_2\). (b) Cavity dressing \(\xi_{\bm k}\) (top panels) and ground-state phase diagrams (bottom panels) obtained without the approximation. Here, \(Q\) is the effective width of the coupling strength~\eqref{sm_coupling_strength}, and \(b_M = 4\pi/(\sqrt{3}a_M)\) is the size of the moir\'e Brillouin zone.}
\end{figure*}

Figure~\ref{sm_figure4} shows the ground-state phase diagrams of the cavity-confined TMD heterobilayer obtained with or without making the above approximation. At finite \(Q\), we can see that \(\xi_{\bm k}\) is slightly modified and the highly frustrated phases such as the zigzag phase become bit narrower accordingly, These modifications are consistent with the expectation that the cavity-mediated long-range hoppings now have an effective cutoff length scale \(\sim1/Q\). 
Nevertheless, as inferred from the figure, these changes are almost negligible, and it is clear that all the results remain qualitatively the same, which justifies the validity of our treatment.

\section{Effect of the modification of the Coulomb interaction\label{sec:app:Coulomb}}
We here discuss possible modification of the Coulomb interaction due to the cavity confinement. We recall that, in the present analysis, the only assumption on the Coulomb interaction is that the interaction strength \(U\) is much larger than the hopping amplitudes of tight-binding orbitals of flat-band electrons \(t_i\). This assumption generally holds true in moir\'e materials due to the band flatness. In particular, the ratio between the magnetic interactions \(J_i\), which determines the magnetic phases of the moir\'e bilayer, is independent of the Coulomb interaction strength \(U\) in the present analysis. Since possible modification of the Coulomb interaction is at most an order of \(O(\eta^2)\) compared to the original value as detailed below, we conclude that it does not affect the discussions in the present work.

In moir\'e materials, one can estimate the strength of the Coulomb interaction \(U\) by the matrix element
\begin{align}
    U \equiv \langle w_0|\hat U|w_0\rangle = \int d{\bm r}\int d{\bm r'} U(\bm r-\bm r') w_0^*(\bm r) w_0(\bm r'),
\end{align}
where \(|w_0\rangle\) is the Wannier orbital constructed from the Bloch states of the moir\'e flat band, and \(U(\bm r)= (e^2/\epsilon){[r^{-1}-(r^2+D^2)^{-1/2}]}\) is the screened Coulomb potential in moir\'e materials~\cite{LM20}. In the present setup of cavity moir\'e materials, the modification of the Coulomb interaction originates from the modification of the Bloch states \(|\psi_{0\bm k}\rangle\) due to the interaction with hyperbolic cavities. Within the perturbation theory, the Bloch states are modified as
\begin{widetext}
\begin{align}
    |\psi_{0\bm k}\rangle \to |\psi_{0\bm k}'\rangle = |\psi_{0\bm k}\rangle|0\rangle + \sum_{n\neq 0}\sum_{\bm q} \frac{1}{\hbar\omega_{\bm q}+\varepsilon_{0\bm k}-\varepsilon_{n\bm k-\bm q}}\frac{eA_{\bm q}}{\hbar}(\bm G_{0n}^{\bm k-\bm q,\bm q}\cdot\bm e_{\bm q})^* |\psi_{n\bm k-\bm q}\rangle|1_{\bm q}\rangle + O(\eta^2),
\end{align} 
where \(|0\rangle\) is the cavity vacuum, and \(|1_{\bm q}\rangle = \hat a_{\bm q}^\dagger|0\rangle\) is the one-phonon-polariton excited state with in-plane momentum \(\bm q\). Thus, the Wannier orbital, the superposition of the Bloch states, is modified as
\begin{align}
    |w_0\rangle &\to |w'_0\rangle = |w_0\rangle|0\rangle + \sum_{\bm q}|w_{\bm q}^{(1)}\rangle|1_{\bm q}\rangle + O(\eta^2),\label{sm_wq1}\\
    |w_{\bm q}^{(1)}\rangle &= \int \frac{v_Md^2\bm k}{(2\pi)^2} \sum_{n\neq 0} \frac{1}{\hbar\omega_{\bm q}+\varepsilon_{0\bm k}-\varepsilon_{n\bm k-\bm q}}\frac{eA_{\bm q}}{\hbar}(\bm G_{0n}^{\bm k-\bm q,\bm q}\cdot\bm e_{\bm q})^* |\psi_{n\bm k-\bm q}\rangle,
\end{align}
where we note that \(|w_{\bm q}^{(1)}\rangle\) is \(O(\eta)\).
Since the second term in the right hand side of Eq.~\eqref{sm_wq1} includes the cavity excited state \(|1_{\bm q}\rangle\), the modification of the Coulomb interaction becomes \(O(U\cdot\eta^2)\) as
\begin{align}
    U \to U' = \frac{\langle w_0'|\hat U|w_0'\rangle}{\langle w_0'|w_0'\rangle} = 
    \frac{\langle w_0|\hat U|w_0\rangle + \sum_{\bm q} \langle w_{\bm q}^{(1)}|\hat U|w_{\bm q}^{(1)}\rangle + O(U\cdot\eta^2)}{\langle w_0|w_0\rangle + \sum_{\bm q}\langle w_{\bm q}^{(1)}| w_{\bm q}^{(1)}\rangle + O(\eta^2)} = \langle w_0|\hat U|w_0\rangle + O(U\cdot\eta^2),  
\end{align}
which proves the statement above.
\end{widetext}

\section{Effect of the polariton loss in the hyperbolic cavity\label{sec:app:loss}}
We here discuss a possible effect of loss of the phonon polaritons in the hyperbolic cavity. In general, the loss rate of hyperbolic polaritons is very low compared to the optical phonon frequency in vdW crystals \(\hbar\omega_{\bm q}=10^2\sim 10^3\) meV. To estimate the strength of the polariton loss, we first recall that the frequency \(\omega_{\bm q}\) (and out-of-plane momentum \(\kappa_{\bm q}\)) of the polariton mode with in-plane momentum \(\bm q\) is determined by the eigenequations of electromagnetic fields as~\cite{YA23}
\begin{align}
    \frac{\bm q^2}{\epsilon_{z}(\omega)} + \frac{\kappa^2}{\epsilon_t(\omega)} &= \frac{\omega^2}{\epsilon_0c^2},\label{eq:app:eigen1}\\
    \tan\left(\frac{\kappa d}{2}\right) &= - \frac{\kappa}{-\epsilon_t(\omega)\sqrt{q^2-\frac{\omega^2}{c^2}}}\label{eq:app:eigen2}.
\end{align}
Here, \(\epsilon_{t(z)}(\omega)\) is the in-plane (out-of-plane) permittivity of the dielectric medium, and \(d\) is the thickness of the vdW slabs. To include the effect of phonon loss in vdW materials, we can use the complex-valued permittivities as~\cite{MEY22}
\begin{align}
    \epsilon_z(\omega) &= \epsilon_{z\infty} \left(1 + \frac{\eta_z^2}{\Omega_z^2 - \omega^2 - i\gamma_z\omega}\right),\\
    \epsilon_t(\omega) &=  \epsilon_{t\infty} \left(1 + \frac{\eta_t^2}{\Omega_t^2 - \omega^2 - i\gamma_t\omega}\right).
\end{align}
We note that an ultralow-loss ratio \(\gamma/\Omega\sim 0.005\) has been achieved in hyperbolic vdW materials, such as hBN~\cite{MEY22}.
As \(\bm q\) is a real-valued variable corresponding to the in-plane momentum between the airgap of the cavity, the out-of-plane momentum \(\kappa\) and the cavity frequency \(\omega\) in the solutions of the eigenequation are complex-valued, reflecting the phonon loss in the dielectric medium.

\begin{widetext}
    \begin{figure*}
        \includegraphics[width = 13cm]{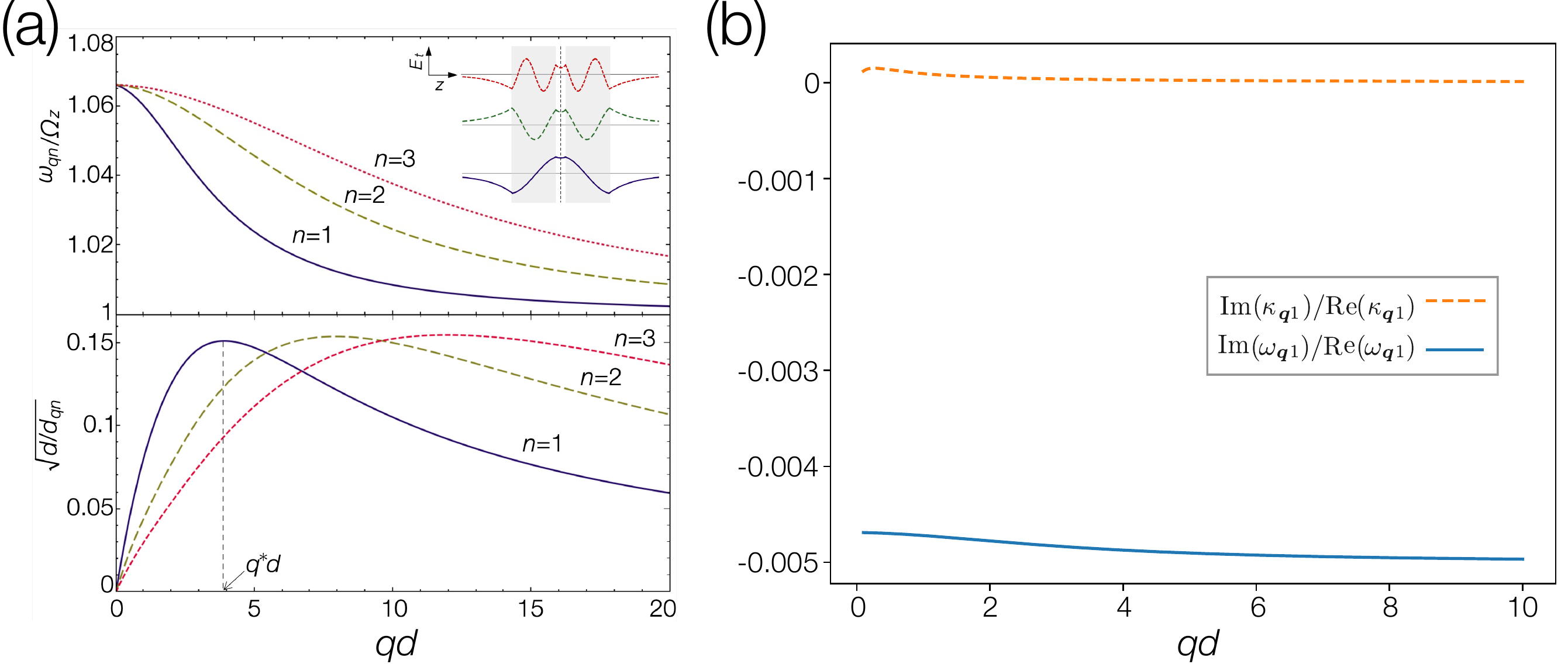}
        \caption{\label{sm_figure5} (a) Dispersions \(\omega_{\bm qn}\) of the hyperbolic phonon polaritons (top panel) and inverse of the square root of the dimensionless effective confinement length \(d_{\bm qn}/d\), which is proportional to the coupling strength \(g_{\bm q}/\omega_{\bm q}\). Reproduced from Fig.~2 in Ref.~\cite{YA23}. (b) The ratio between the imaginary part and the real part of the out-of-plane momenta \(\kappa_{\bm q}\) and the phonon-polariton frequencies \(\omega_{\bm q}\). We plot the ratio for the \(n=1\) eigenmodes (see top panel in (a)). The parameters are the same as in Ref.~\cite{YA23}, and we set the loss ratio as \(\gamma_z/\Omega_z = \gamma_t/\Omega_t = 0.01\).}
    \end{figure*}
\end{widetext}

Figure~\ref{sm_figure5} shows the ratio between the imaginary part and the real part of \(\kappa_{\bm q}\) and \(\omega_{\bm q}\), which are obtained by solving the eigenequations~\eqref{eq:app:eigen1}, \eqref{eq:app:eigen2} up to the first order of \(\gamma\). 
The changes of \(\kappa_{\bm q}\) and  \(\omega_{\bm q}\) could in principle affect the value of the coupling strength; for instance, the change of  \(\kappa_{\bm q}\)  modifies the spatial profile of each eigenmode and its mode amplitude. However, from the results we have obtained, we estimate the amount of the change in the coupling strength to be $\sim$0.5\%, which only leads to the negligibly small shifts of the proposed phase boundaries. We thus expect that the polariton loss does not play an important role in our consideration based on the setup consisting of hyperbolic cavity with ultralow phonon losses.

\clearpage
\bibliography{masuki_bib}

\end{document}